\documentclass[useAMS,usenatbib]{mn2e}

\usepackage{aas_macros}
\usepackage{graphicx}
\usepackage{amsmath}
\usepackage{xspace}
\usepackage{ifthen}
\usepackage{lscape}
\usepackage{placeins}
\usepackage{longtable}

\usepackage{array}
\newcolumntype{L}[1]{>{\raggedright\let\newline\\\arraybackslash\hspace{0pt}}p{#1}}
\newcolumntype{C}[1]{>{\centering\let\newline\\\arraybackslash\hspace{0pt}}p{#1}}
\newcolumntype{R}[1]{>{\raggedleft\let\newline\\\arraybackslash\hspace{0pt}}p{#1}}

{ 

\title[Secondary eclipses from the AAT]{Secondary eclipse observations for seven hot-Jupiters from the Anglo-Australian Telescope
\thanks{Based on observations obtained at the Anglo-Australian Telescope, Siding
Spring, Australia.}}

\author[G.~Zhou et al.]
{\parbox{\textwidth}
{G.~Zhou$^{1}$\thanks{E-mail: \texttt{george.zhou@anu.edu.au}},
D.D.R.~Bayliss$^{2,1}$,
L.~Kedziora-Chudczer$^{3,4}$,
C.G.~Tinney$^{3,4}$,
J.~Bailey$^{3,4}$,
G.~Salter$^{5}$,
and J.~Rodriguez$^{6}$
\vspace{0.4cm}}\\
\parbox{\textwidth}{
$^{1}${Research School of Astronomy and Astrophysics, Australian National University, Canberra, ACT 2611, Australia}\\
$^{2}${Observatoire Astronomique de l'Universit\'{e} de Gen\`{e}ve, 51 ch. des Maillettes, 1290 Versoix, Switzerland}\\
$^{3}${School of Physics, University of New South Wales, Sydney, NSW 2052, Australia}\\
$^{4}${Australian Centre for Astrobiology, University of New South Wales, Sydney, NSW 2052, Australia}\\
$^{5}${Laboratoire d'astrophysique de Marseille, Technopôle de Marseille-Etoile 38, rue Frédéric Joliot-Curie, 13388 Marseille cedex 13, France}\\
$^{6}${Department of Physics and Astronomy, Vanderbilt University, 6301 Stevenson Center, Nashville, TN 37235, USA}\\
}}

\begin{document}

\date{Accepted 2015 September 14. }

\pagerange{\pageref{firstpage}--\pageref{lastpage}} \pubyear{2015}

\maketitle

\label{firstpage}

\begin{abstract}
We report detections and constraints for the near infrared $Ks$ band secondary eclipses of seven hot-Jupiters using the IRIS2 infrared camera on the Anglo-Australian Telescope. Eclipses in the $Ks$ band for WASP-18b and WASP-36b have been measured for the first time. We also present new measurements for the eclipses of WASP-4b, WASP-5b, and WASP-46b, as well as upper limits for the eclipse depths of WASP-2b and WASP-76b. In particular, two full eclipses of WASP-46b were observed, allowing us to demonstrate the repeatability of our observations via independent analyses on each eclipse. Significant numbers of eclipse depths for hot-Jupiters have now been measured in both $Ks$ and the four \emph{Spitzer} IRAC bandpasses. We discuss these measurements in the context of the broadband colours and brightness temperatures of the hot-Jupiter atmosphere distribution. Specifically, we re-examine the proposed temperature dichotomy between the most irradiated, and mildly irradiated planets. We find no evidence for multiple clusters in the brightness temperature – equilibrium temperature distributions in any of these bandpasses, suggesting a continuous distribution of heat re-emission and circulation characteristics for these planets.

\end{abstract}


\section{Introduction}
\label{sec:introduction}

A secondary eclipse occurs when the emergent flux from a planet is blocked by its host star, which allows the direct measurement of the infrared day side temperature of a hot-Jupiter. Assembling a sample of secondary eclipses is one way to comparatively study the atmospheres of the hot-Jupiter population. For example, \citet{2011ApJ...729...54C} have compared measured effective temperatures to the incident stellar irradiation received by hot-Jupiters, and proposed that the most irradiated planets have heat re-emission properties that are different from those for cooler planets. Updated results from \citet{2015arXiv150206970S} have constructed ensemble emission spectra of the hot-Jupiter population, and placed further constraints on the albedo and heat recirculation efficiences of the most well characterised hot-Jupiters. \citet{2014MNRAS.439L..61T} and \citet{2014MNRAS.444..711T} have used multi-band eclipse measurements to create colour-magnitudes diagrams of hot-Jupiters, enabling an empirical comparison with the broadband spectral features of M-dwarfs and brown dwarfs. The presence of proposed atmospheric thermal inversion features have also been linked to stellar activity \citep{2010ApJ...720.1569K} via comparative studies. 

The majority of secondary eclipse measurements have been made using the \emph{Spitzer Space Telescope} in the IRAC bands \citep[e.g.][]{2014arXiv1411.7404D,2014ApJ...788...92S,2014A&amp;A...572A..73L}. Forty eight planets have now been sampled in eclipse at the IRAC $4.5\,\mu\text{m}$ band. In comparison, only 26 planets have been observed in the $Ks$ band from the ground.  However, ground-based eclipse observations can probe shorter wavelengths that are inaccessible to \emph{Spitzer}, and so probe deeper into the planetary atmosphere and examine different regimes of atmospheric circulation. They also provide an extended wavelength baseline for comparison with spectral models. Despite the challenges inherent in ground-based eclipse measurements for hot-Jupiters, a number of facilities are now consistently delivering eclipse depth measurements. These include the CFHT \citep{2010ApJ...717.1084C,2010ApJ...718..920C,2011AJ....141...30C,2013ApJ...770...70W,2015ApJ...802...28C}, 200-inch at Palomar \citep[e.g.][]{2012ApJ...744..122Z,2012ApJ...748L...8Z,2014ApJ...781..109O,2014ApJ...788...92S,2014ApJ...796..115Z}, and the ESO 2.2-m telescope \citep[e.g.][]{2014A&amp;A...563A..40C,2014A&amp;A...564A...6C,2014A&amp;A...567A...8C}.

In \citet{2014MNRAS.445.2746Z}, we introduced the series of eclipse observations we are performing at the Anglo-Australian Telescope (AAT), with the aim of measuring $Ks$ band eclipses for a large number of hot-Jupiters in the southern hemisphere. In this paper, we report eclipse measurements and constraints for seven hot-Jupiters: WASP-2b, -4b, -5b, -18b, -36b, -46b, and -76b. The observations and analysis are described in Section~\ref{sec:observations}, results and comparisons between previous observations reported in Section~\ref{sec:results}. Section~\ref{sec:discussion} discusses the eclipse observations in the context of the hot-Jupiter colour-magnitude, colour-colour diagrams, and brightness temperature -- equilibrium temperature relationships. 

\section{Observations and analysis}
\label{sec:observations}

\subsection{Observing strategy and data reduction}
\label{sec:observ-data-reduct}

These eclipse observations were performed using the IRIS2 instrument \citep{2004SPIE.5492..998T} on the 3.9-m AAT at Siding Spring Observatory, Australia. IRIS2 is a $1\text{K}\times1\text{K}$ infrared camera with a HAWAII-1 HgCdTe infrared detector, read out over four quadrants in double-read mode. The instrument has a field of view of $7'.7 \times 7'.7$ and plate scale of 0.4486''/pixel. The observing strategy for each eclipse observation is similar to that described in \citet{2014MNRAS.445.2746Z}: the telescope is defocused to broaden the stellar point-spread function, reducing the effect of intra- and inter-pixel systematics, and preventing saturation of the target and key reference stars. Exposure times are set such that the target and key reference stars are kept below peak counts of 20,000 ADU, so as to keep within the regime where detector non-linearity is minimised (non-linearity of $>1$\% occurs above 40,000 ADU). We apply a non-linearity correction to each image, following the IRIS2 manual's prescription, to correct the very small ($<0.05\%$) non-linearity present below 20,000 ADU. Typical exposure times are $<10$\,s per exposure. The WASP-18 observations were taken using 5s exposures, with 20 exposures being averaged and saved as a single frame. This observing mode was tested to reduced the data volume of the observations. However we found that it led to a reduction in the precision of eclipse timing and depth measurements, and it was not used for any subsequent observations. For all other observations, single exposures are saved and used in the analysis. Observations of six offset positions are taken before and after each eclipse sequence to sample the sky background. These observations are used as flat fields in the reduction process. The field is carefully centred such that the target and key reference stars do not fall on bad pixels. The eclipse sequence is performed in stare mode, with the telescope guided to minimise drift of the field. Dark frames of the same exposure time as the eclipse observations are taken before and after each night. Tests of darks taken through an experimental night showed no drifts in the dark current. Information on the systems observed, and the specifics of each eclipse observation, including date, number of exposures, median cadence, median point-spread function full width at half maximum (FWHM), are given in Table~\ref{tab:observations_list}. 

\begin{table*}
  \centering
  \caption{Targets and observation details}
  \label{tab:observations_list}
  \begin{tabular}{lrlrrrr}
    \hline\hline
    Target & $K$mag$^a$ & Observation& Number of & Cadence (s) & Median & No. Ref\\
    & & Date \& Time (UT)$^b$ & Exposures & &FWHM (pix) & Stars\\

    \hline
    WASP-2 & 9.6 & 2014-09-10 08:54--14:46 & 3100 & 6 & 7.6 & 9\\
    WASP-4 & 10.7 & 2014-09-04 15:54--17:02 & 354 & 11 & 17.4 & 4\\
    & & 2014-09-11 09:23--11:38 & 689 & 11 & 14.2 & 4\\
    WASP-5 & 10.6 & 2014-09-14 13:17--19:17 & 1727 & 11 & 12.3 & 6\\
    WASP-18 & 8.1 & 2014-09-05 11:16--14:54 & 95 & 120$^c$ & 19.0 & 3\\
    WASP-36 & 11.3 & 2015-03-09 10:21--15:51 & 4358 & 4 & 4.1 & 9\\
    WASP-46 & 11.4 & 2014-09-11 12:07--17:28 & 1600 & 11 & 13.4 & 6\\
    & & 2014-09-14 08:55--12:53 & 1230 & 11 & 11.6 & 6\\
    WASP-76 & 8.2 & 2014-09-13 13:26--19:16 & 2953 & 6 & 17.4 & 2 \\
    \hline
  \end{tabular}
\begin{flushleft} 
$^a$ 2MASS magnitudes \citep{2006AJ....131.1163S}.\\
$^b$Start and end of each observation sequence.\\
$^c$$20\times5$s exposures are averaged and used for analysis. Individual exposures were not saved.  
\end{flushleft}
\end{table*}

Each object frame is dark subtracted and flat divided. A master dark frame is median combined from darks of the same exposure time, taken on the same night as the object frames. Master flat fields are created from the set of offset frames taken before and after the eclipse sequence, with stars masked, and median combined. Bad pixels in the object frames are then interpolated over using the surrounding pixels via a radial basis function interpolation. Baryocentric Julian Date (BJD) time stamps for each frame are calculated using the converttime task in VARTOOLS \citep{2008ApJ...675.1233H,2010PASP..122..935E}.

For each frame, stars are identified using Source Extractor \citep{1996A&amp;AS..117..393B} and cross matched using \emph{grmatch} task in FITSH \citep{2012MNRAS.421.1825P}. Coordinates for the target and reference stars are transformed using \emph{grtrans}, and aperture photometry performed using \emph{fiphot}. For each set of observations, we extract the photometry through a set of fixed apertures. Background flux beneath each aperture is measured in a 5 pixel wide annulus around each central aperture. Adjacent stars within this annular background aperture are masked before the background is calculated. We also tried extractions using variable apertures, by setting the aperture size per frame as a multiple of the average FWHM of the stellar point-spread functions. However, extracting photometry using fixed apertures yielded light curves with the least out-of-eclipse scatter. We conclude photometry using variable aperture sizes introduces noise associated with the FWHM estimate, and especially for fields with just a few bright reference stars.

\subsection{Light curve analysis and modelling}
\label{sec:light-curve-analysis}

The object light curve is corrected by a master reference light curve, constructed from selected reference stars. Weights are applied to each reference star light curve such that the out-of-eclipse scatter of the object light curve is minimised after correction. The final light curve is found by minimising the out-of-transit scatter of the object light curve for all permutations of object and reference star extraction apertures. 

The eclipse light curves are fitted with the \citet{1972ApJ...174..617N} model, using an adapted implementation of the JKTEBOP code \citep{1981AJ.....86..102P, 2004MNRAS.351.1277S}. The eclipses are modelled with free parameters $e\cos\omega$, which determines the phase of the eclipse, and the surface brightness ratio $S_p/S_\star$, which determines the depth of the eclipse. To propagate the uncertainties in the system parameters, we also incorporate the free parameters period $P$, primary transit reference time $T_0$, planet-star radius ratio $R_p/R_\star$, normalised orbit radius $(R_p+R_\star)/a$, and line-of-sight inclination $i$, each constrained tightly by Gaussian priors adapted from the literature uncertainties. In the cases where no clear eclipse is seen, we constrain the $e\cos\omega$ with Gaussian priors using literature eccentricity values from previous \emph{Spitzer} eclipse measurements, or radial velocity constraints. The $e\cos\omega$ constraints, when applied, are noted in Table~\ref{tab:sys_param}. The light travel time has been accounted for when fitting for the eclipse timing. 

To incorporate the influence of instrumental and atmospheric variations into the reported uncertainties, we model the light curve as a function of external parameters simultaneously with the model fit. This is even more important for infrared light curves than at optical wavelengths, given the greater pixel-to-pixel sensitivity variations in infrared detectors, combined with the much larger variability of the infrared sky background. The influence of the external parameters is modelled as a linear combination of factors which include \citep[as per][]{2014MNRAS.445.2746Z}: time $t$, target star pixel positions $X$ and $Y$, stellar point-spread functions FWHM $F$, background counts $B$, and airmass $A$. Fits are performed using all combinations of the external parameters. We then adopt the model with the set of external parameters that minimise the Bayesean Information Criterion (BIC) post-fitting. The external parameter model components adopted for each eclipse observation are listed in Table~\ref{tab:sys_param}. We note in Section~\ref{sec:results} when the next-best decorrelation models, ranked by BIC, exhibit a different eclipse fit. In the cases where eclipse observations are combined from multiple nights (WASP-4b and WASP-46b), the external parameters and coefficients are independent for each night. This allows us to account for the different factors that affect the observations each time. For example, the seeing conditions were stable for the WASP-46b observation on 2014-09-11, and variable for 2014-09-14, the instrument model thus contained only the time and airmass components for the first night, and time, airmass, and FWHM components for the second.

The best fit model parameters and associated uncertainties are derived using an Markov chain Monte Carlo (MCMC) analysis, via the \emph{emcee} ensemble sampler \citep{2013PASP..125..306F}. The MCMC analyses are run twice, with the walkers for the second run originating from the best fit parameters from the first run. In the second run, the per-point photon errors are inflated to force a reduced $\chi^2=1$. This allows error sources other than photon-noise to be included in the uncertainty estimate, and is particularly important for light curves with substantial red noise components. 

\section{Results}
\label{sec:results}

\begin{figure*}
  \centering
  \includegraphics[width=13cm]{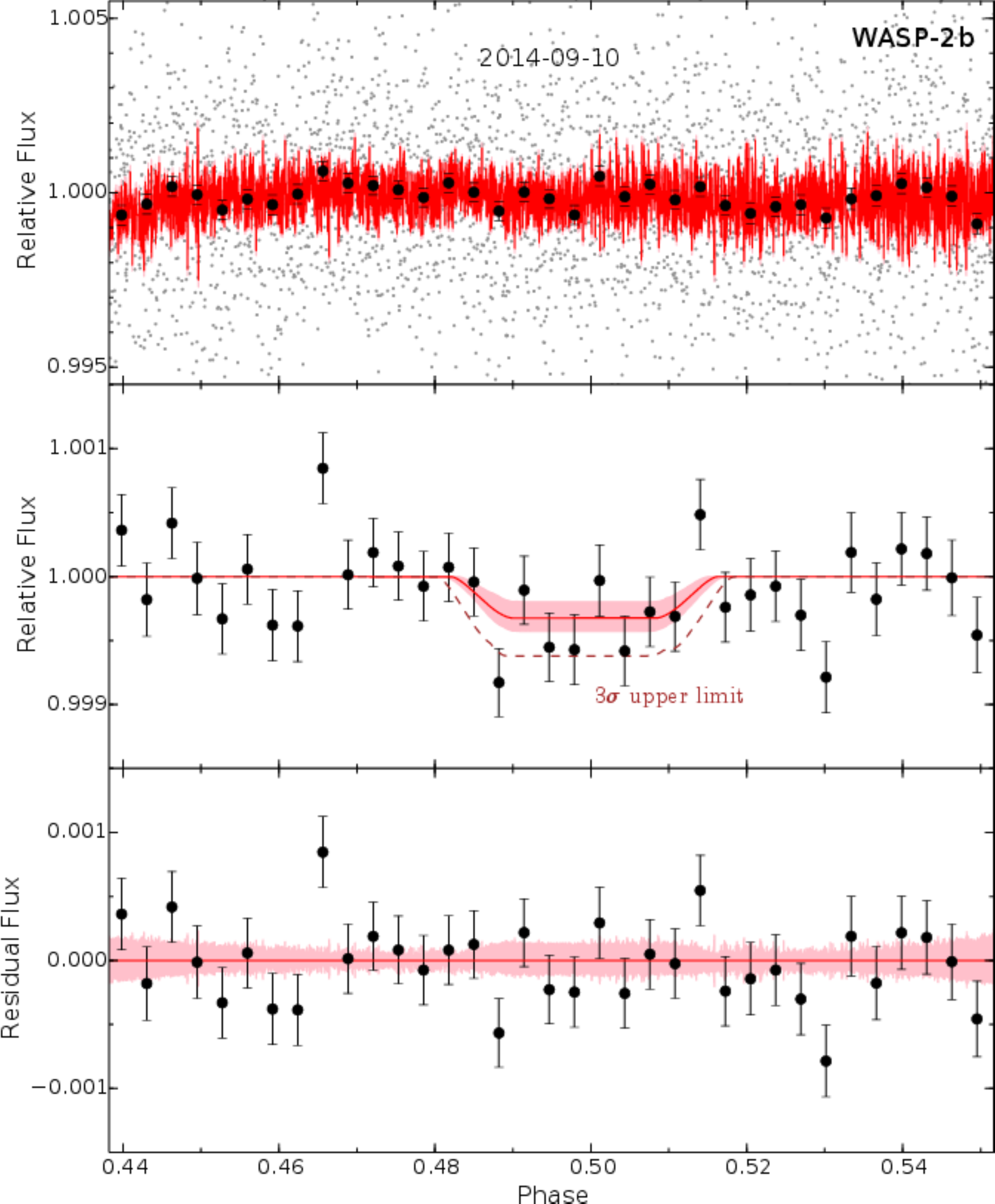}\\
  \caption{\textbf{Top}: The $Ks$ band relative photometry light curve for the eclipse event of WASP-2b as measured on 2014-09-10. Photometry from individual observations are plotted in gray, 10 minute bins in black. For each bin containing $n$ points, the error bars are plotted to represent the mean per-point uncertainties, which are photon errors inflated to force a reduced $\chi^2=1$, scaled by $1/\sqrt{n}$. The best fit model (eclipse and instrumental parameters) is plotted in red, and 68\% of the allowed models reside within the shaded pink region. \textbf{Middle}: As above, but with only the 10 minute binned data and the instrument model subtracted. The shaded regions show the allowed models in terms of the transit parameters only. Since we can only place an upper limit on the eclipse of WASP-2b, the dashed brown line represents the $3\sigma$ regime of the models. \textbf{Bottom}: As above, but showing the data residuals to the best fit model. The shaded region shows the difference between each allowed model (eclipse and instrumental) to the best fit model.}
  \label{fig:eclipse_lc_WASP2}
\end{figure*}

\begin{figure*}
  \centering
  \includegraphics[width=13cm]{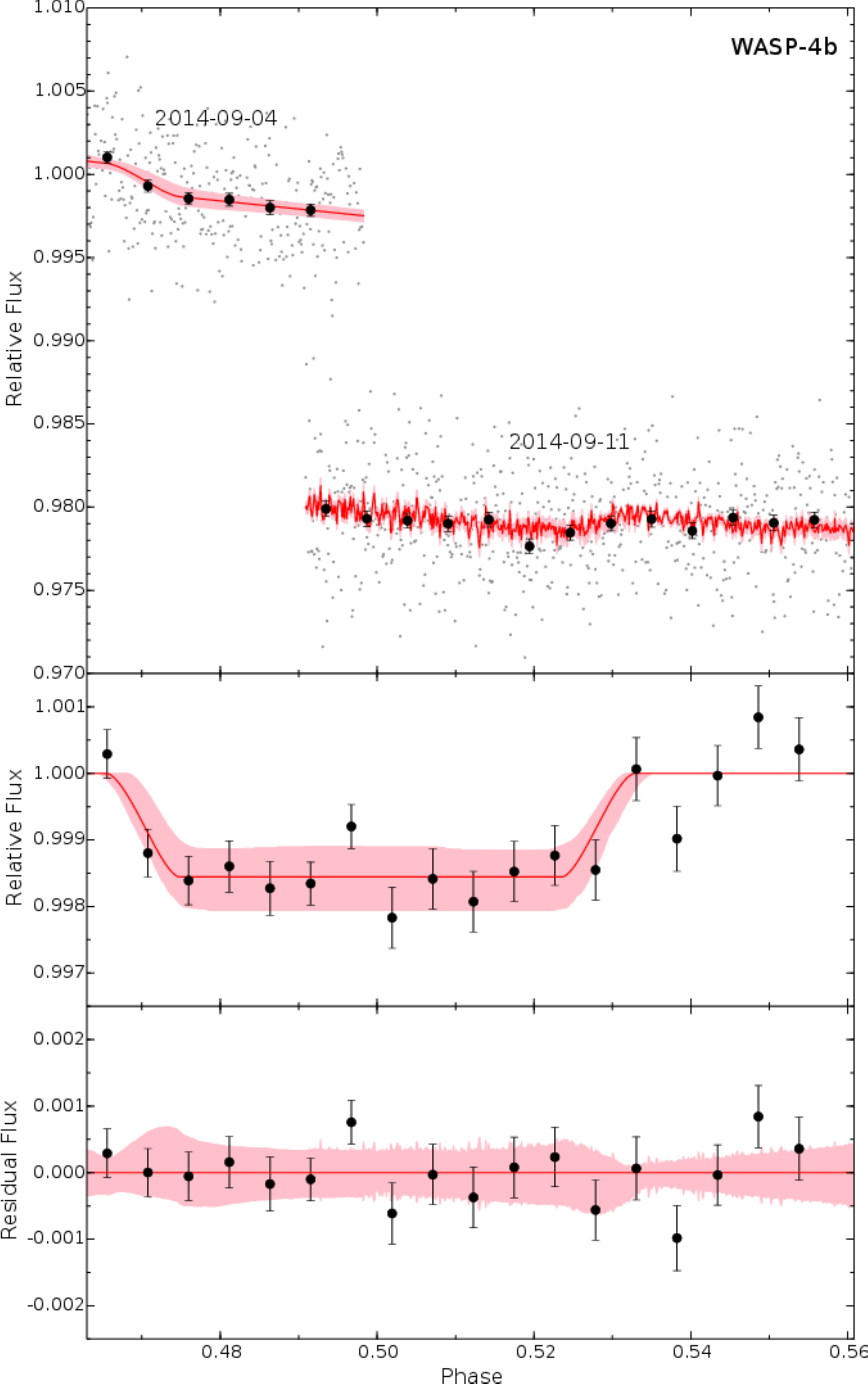}\\
  \caption{Light curves for the eclipse events of WASP-4b as observed on 2014-09-04 and 2014-09-11, labelled as per Figure~\ref{fig:eclipse_lc_WASP2}, but with the light curves from each night arbitrary offset for clarity in the top panel.}
  \label{fig:eclipse_lc_WASP4}
\end{figure*}

\begin{figure*}
  \centering
  \includegraphics[width=13cm]{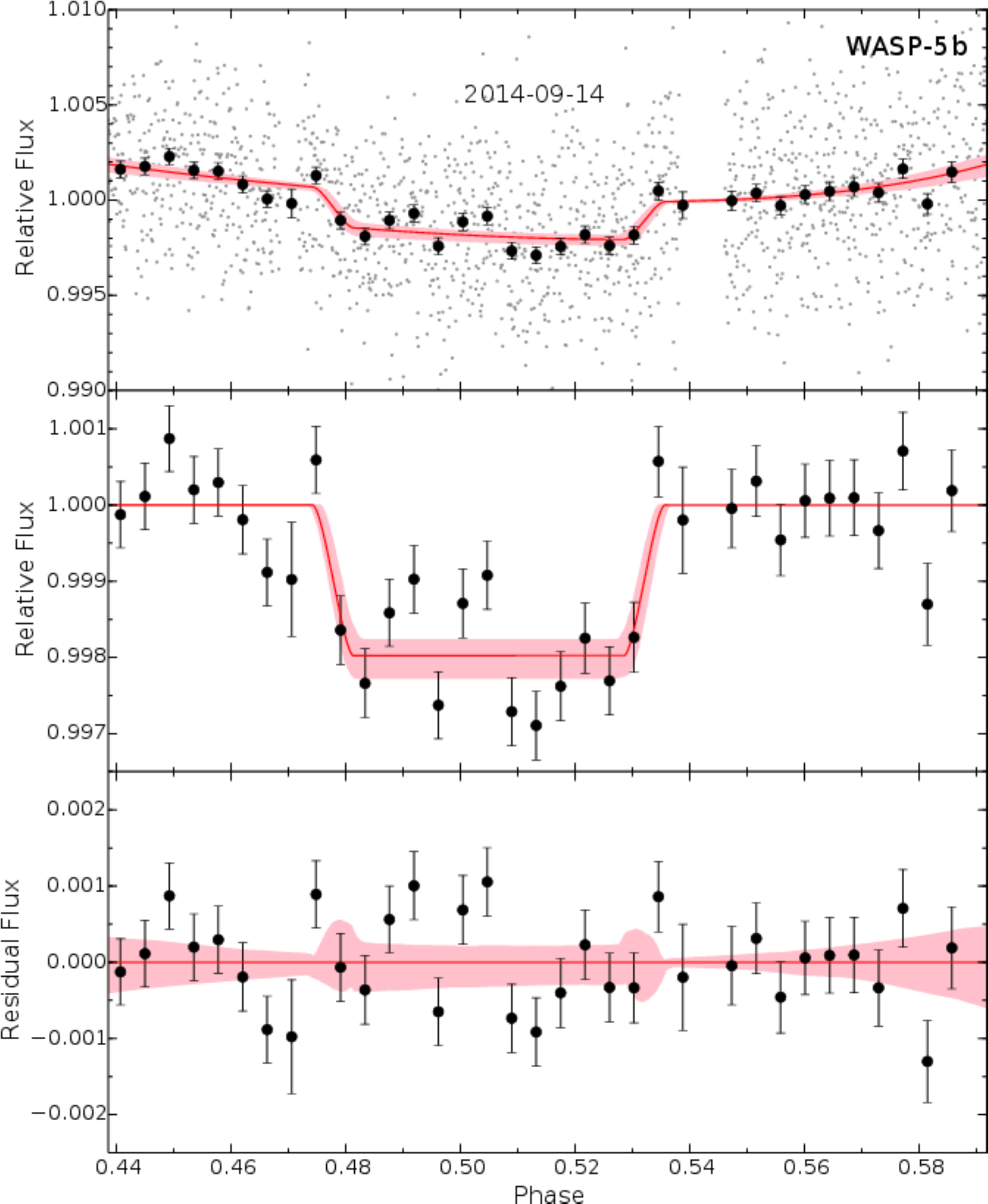}\\
  \caption{Light curve for the eclipse event of WASP-5b, plotted as per the description for Figure~\ref{fig:eclipse_lc_WASP2}.}
  \label{fig:eclipse_lc_WASP5}
\end{figure*}

\begin{figure*}
  \centering
  \includegraphics[width=13cm]{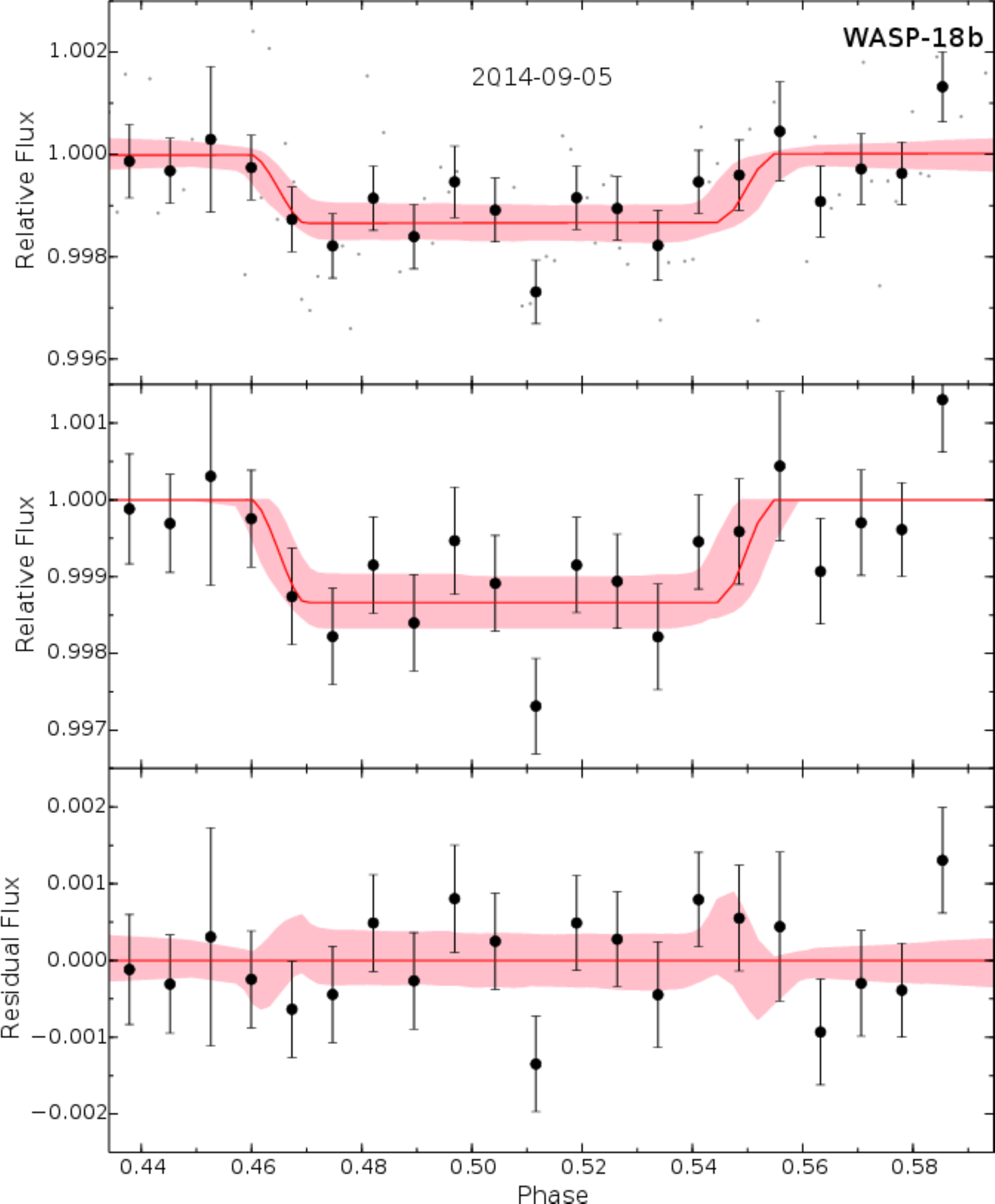}\\
  \caption{Light curve for the eclipse event of WASP-18b, plotted as per the description for Figure~\ref{fig:eclipse_lc_WASP2}.}
  \label{fig:eclipse_lc_WASP18}
\end{figure*}

\begin{figure*}
  \centering
  \includegraphics[width=13cm]{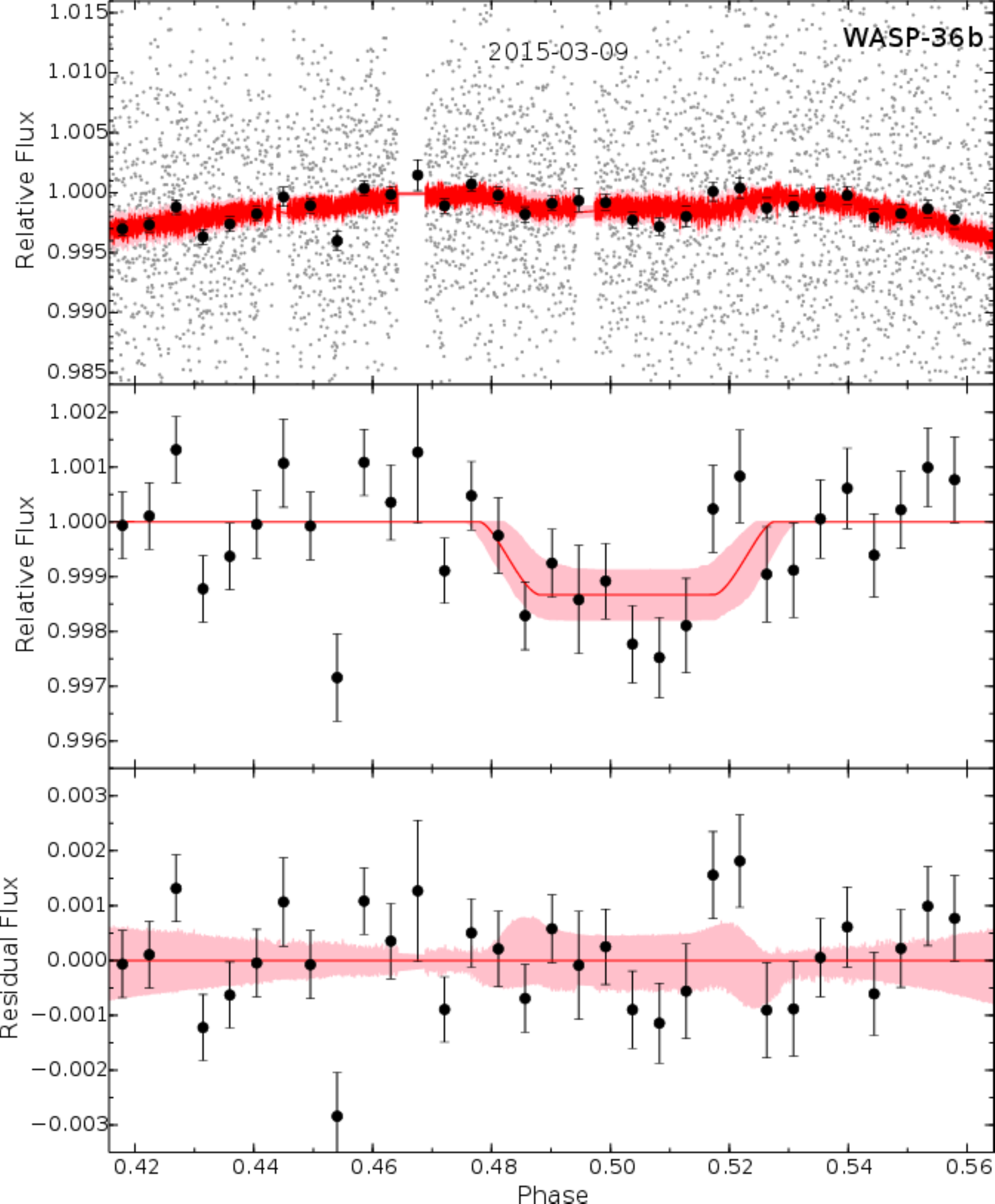}\\
  \caption{Light curve for the eclipse event of WASP-36b, plotted as per the description for Figure~\ref{fig:eclipse_lc_WASP2}.}
  \label{fig:eclipse_lc_WASP36}
\end{figure*}

\begin{figure*}
  \centering
  \includegraphics[width=13cm]{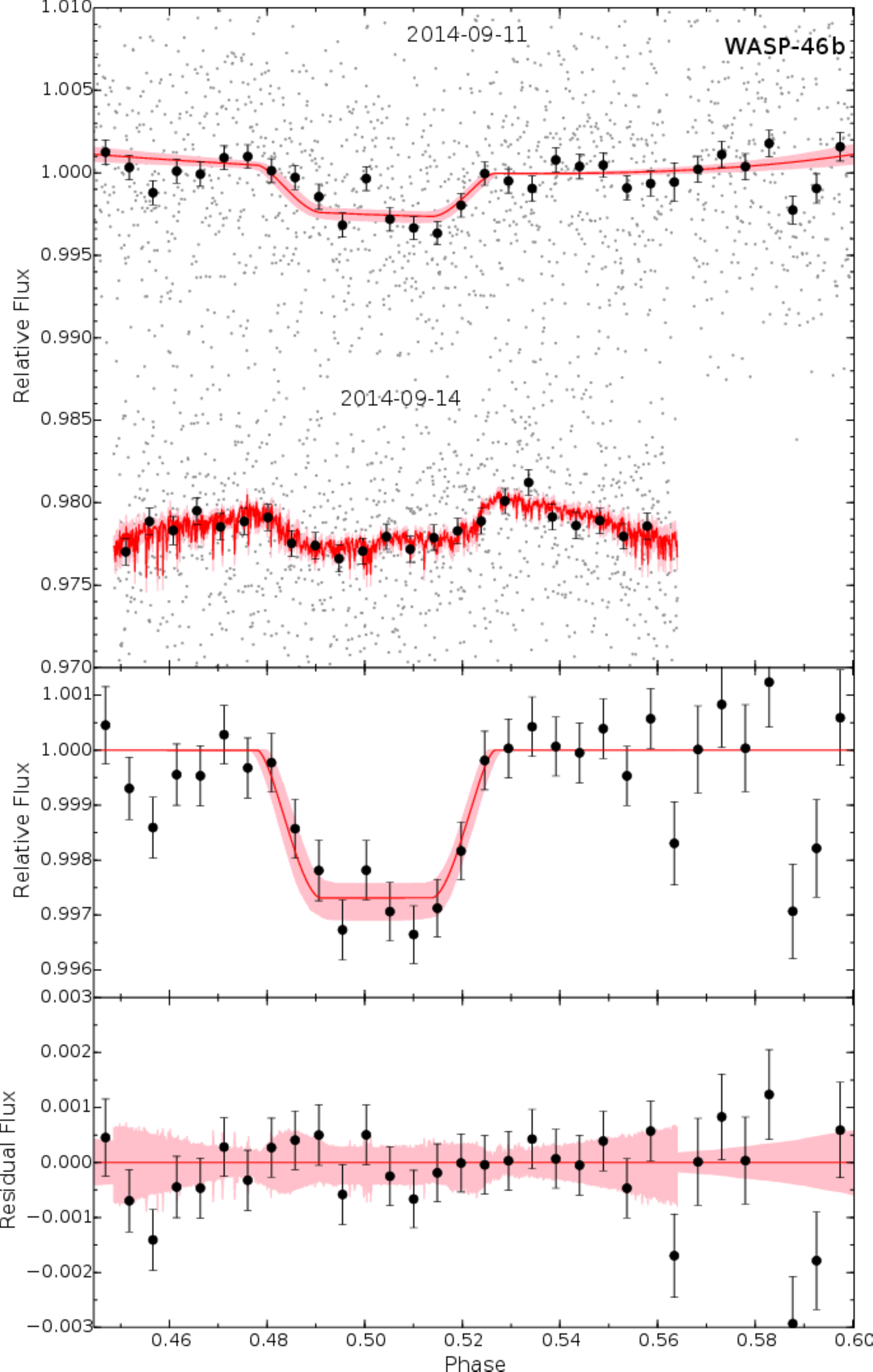}\\
  \caption{Light curves for the eclipse events of WASP-46b, as measured from 2014-09-11 and 2014-09-14, plotted as per the description for Figure~\ref{fig:eclipse_lc_WASP4}.}
  \label{fig:eclipse_lc_WASP46}
\end{figure*}

\begin{figure*}
  \centering
  \includegraphics[width=13cm]{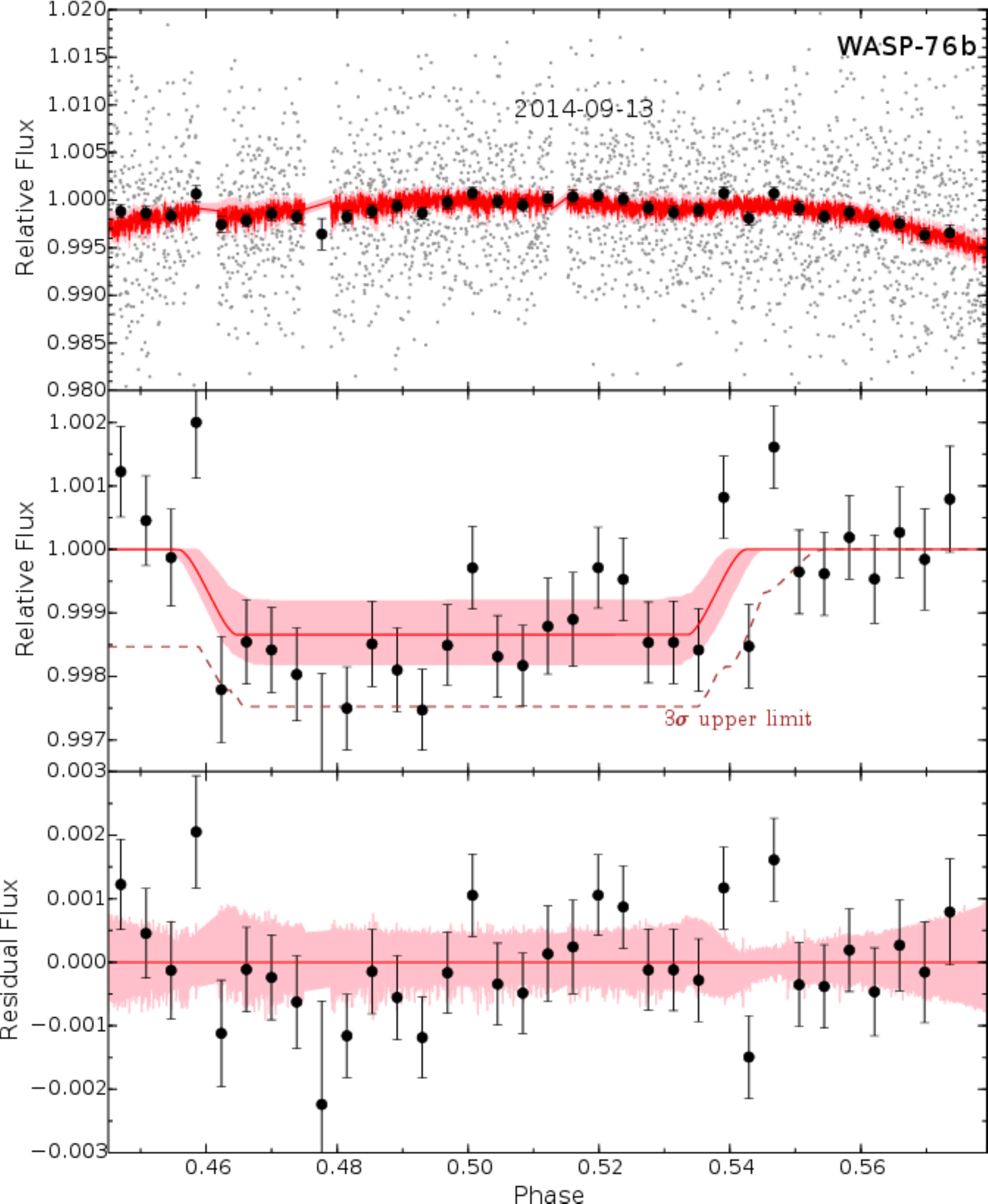}\\
  \caption{Light curve for the eclipse event of WASP-76b, plotted as per the description for Figure~\ref{fig:eclipse_lc_WASP2}.}
  \label{fig:eclipse_lc_WASP76}
\end{figure*}

We detect the eclipses of WASP-4b, -5b, -18b, -36b, and -46b at $>3\sigma$ significance, and provide the $3\sigma$ upper limits for the eclipses of WASP-2b and -76b. The full set of derived parameters, including flux ratios, eccentricity constraints, and brightness temperatures, are listed in Table~\ref{tab:sys_param}. The light curves for the eclipse observations are plotted in  Figures~\ref{fig:eclipse_lc_WASP2}--\ref{fig:eclipse_lc_WASP76}, with the $3\sigma$ upper limits marked where appropriate. 

For WASP-2b, we determine a $3\sigma$ upper limit to the eclipse depth of $<0.07$\% from a single eclipse observation. The eclipse of WASP-2b has previously been measured with \emph{Spitzer} \citep{2010arXiv1004.0836W} at 3.6, 4.5, 5.9, and $8.0\,\mu \text{m}$. Given the lower flux ratio expected in the $Ks$ band compared to the \emph{Spitzer} bands, and the shallow $0.083\pm0.035$\% eclipse at $3.6\,\mu\text{m}$, our upper limit is consistent with the previous observations. The eclipse phase in our model fit was constrained by a Gaussian prior centred on the \emph{Spitzer} eclipse detection. The AAT-IRIS2 light curve for the eclipse event of WASP-2b is plotted in Figure~\ref{fig:eclipse_lc_WASP2}. 

For WASP-4b, we use two partial eclipses to determine an eclipse depth of $0.16_{-0.04}^{+0.04}$\%, and phase consistent with circular orbit of $e\cos\omega=-0.001_{-0.003}^{+0.003}$. The $Ks$ band eclipse has been previously measured by \citet{2011A&amp;A...530A...5C}, using ISAAC on the VLT, at a depth of $0.185_{-0.013}^{+0.014}$\%, consistent with our measurement to within $1\sigma$. The eclipses were also measured by \emph{Spitzer} at  $3.6\, \mu \text{m}$ and $4.5\, \mu \text{m}$ \citep{2011ApJ...727...23B}. The eclipse phase we measure is also consistent to $1\sigma$ with that measured by \citet{2011A&amp;A...530A...5C} and \citet{2011ApJ...727...23B}. The AAT-IRIS2 light curves for the eclipse events of WASP-4b are plotted in Figure~\ref{fig:eclipse_lc_WASP4}. 

For WASP-5b, we observe an eclipse with a depth of $0.20_{-0.02}^{+0.02}$\%, and an eccentricity estimate from the eclipse phase of $e\cos\omega = 0.008_{-0.002}^{+0.002}$. A $Ks$ band eclipse depth of $0.269_{-0.062}^{+0.062}$\% was measured by \citet{2014A&amp;A...564A...6C} using GROND on the MPG 2.2m telescope. The eclipse has also been measured in the $J$ band \citep{2014A&amp;A...564A...6C}, and at the 3.6 and $4.5\,\mu\text{m}$ \emph{Spitzer} bands \citep{2013ApJ...773..124B}. Our $e\cos\omega$ measurement indicates the orbit is eccentric with a statistical signficiance of $4\sigma$, and is consistent within $1\sigma$ with \citet{2014A&amp;A...564A...6C}, and within $2\sigma$ with \citet{2013ApJ...773..124B}. WASP-5b has a short period, leading to a short tidal circularisatin timescale. Following Equation 1 of \citet{2004ApJ...610..464D}, the tidal circularisation timescale should be $\sim 1$ Myr (assuming a tidal quality factor of $Q'=10^5$). It is interesting that the eccentricty is non-zero, suggesting that perhaps a larger tidal quality factor is required to describe the system. The AAT-IRIS2 light curve for the eclipse event of WASP-5b is plotted in Figure~\ref{fig:eclipse_lc_WASP5}. 

For WASP-18b, we measure an eclipse depth of $0.14_{-0.03}^{+0.03}$\%, and an eccentricity estimate from the eclipse phase of $e\cos\omega = 0.012_{-0.008}^{+0.007}$. The eclipses have previously been measured by \emph{Spitzer} at 3.6 and $4.5\,\mu \text{m}$ by \citet{2011ApJ...742...35N,2013MNRAS.428.2645M}, and at 5.9 and $8.0\,\mu \text{m}$ by \citet{2011ApJ...742...35N}. The eclipse phase we derive is consistent (at the $2\sigma$ level) with that expected from circular orbit, and the measurements from the \emph{Spitzer} observations. The long cadence observing strategy used for the WASP-18b observations results in a larger uncertainty in eclipse phase than for other targets. We also caution that the second-best model, ranked by $\Delta$ BIC (with $\Delta$ BIC of 4), involves the time and background flux terms, reducing the surface flux ratio to $S_p/S_\star = 0.10 _{-0.06}^{+0.07}$. The third-best detrending model involves the detector $y$-position, and gave an eclipse depth consistent with that of the best fit model. The AAT-IRIS2 light curve for the eclipse event of WASP-18b is plotted in Figure~\ref{fig:eclipse_lc_WASP18}. 

For WASP-36b, we report the first eclipse detection of the system at depth of $0.13_{-0.04}^{+0.04}$\%, and eccentricity of $0.004_{-0.005}^{+0.006}$, consistent with a circular orbit. No previous eclipse observations have been reported for this planet. The AAT-IRIS2 light curve for the eclipse event of WASP-36b are plotted in Figure~\ref{fig:eclipse_lc_WASP36}. 

For WASP-46b, we measure an eclipse depth of $0.26_{-0.03}^{+0.05}$\%, and $e\cos\omega = 0.004_{-0.004}^{+0.004}$, as measured from two full eclipse observations. The $Ks$ band eclipse has previously been measured by \citet{2014A&amp;A...567A...8C}, with a reported depth of $0.253_{-0.060}^{+0.063}$\%, consistent with our measurement to better than $1\sigma$. \citet{2014A&amp;A...567A...8C} also measured the $J$ and $H$ band eclipses using GROND on the ESO 2.2m. The eclipse timing is also consistent with that of a circular orbit, and that reported by \citet{2014A&amp;A...567A...8C} to $1\sigma$. The AAT-IRIS2 light curves for the eclipse events of WASP-46b are plotted in Figure~\ref{fig:eclipse_lc_WASP46}.

For WASP-76b, we find a $3\sigma$ upper limit eclipse depth of 0.3\%, with a marginal detection at $2.3\sigma$ of a $0.13_{-0.06}^{+0.06}$\% eclipse. The eclipse phase fit was constrained by a Gaussian prior on the orbit eccentricity \citep{2013arXiv1310.5607W}. The marginal eclipse detection is not well constrained in eclipse phase due to the lack of sufficient pre-ingress baseline. As a result, the $3 \sigma$ upper limit is asymmetric about the eclipse centre. No previous eclipse observations have been reported for this planet. The AAT-IRIS2 light curve for the eclipse event of WASP-76b is plotted in Figure~\ref{fig:eclipse_lc_WASP76}.

\begin{figure*}
  \centering
  \begin{tabular}{ccc}
    \includegraphics[width=5cm]{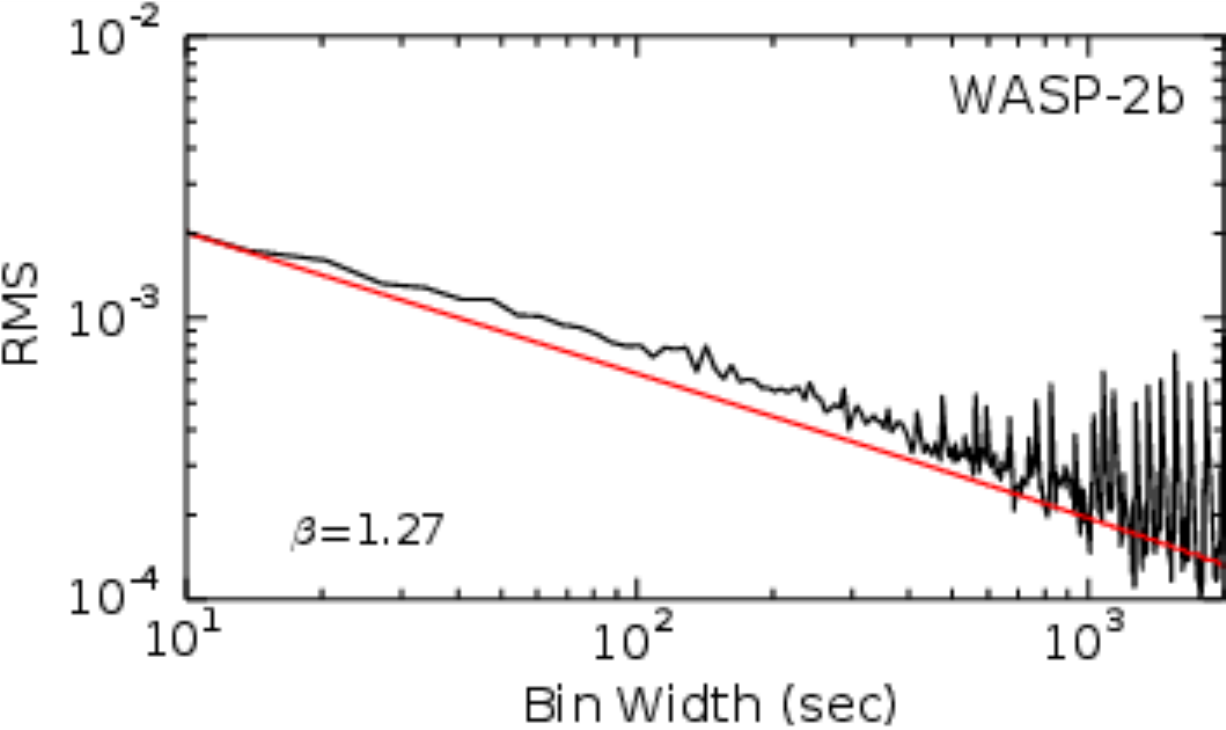} &
    \includegraphics[width=5cm]{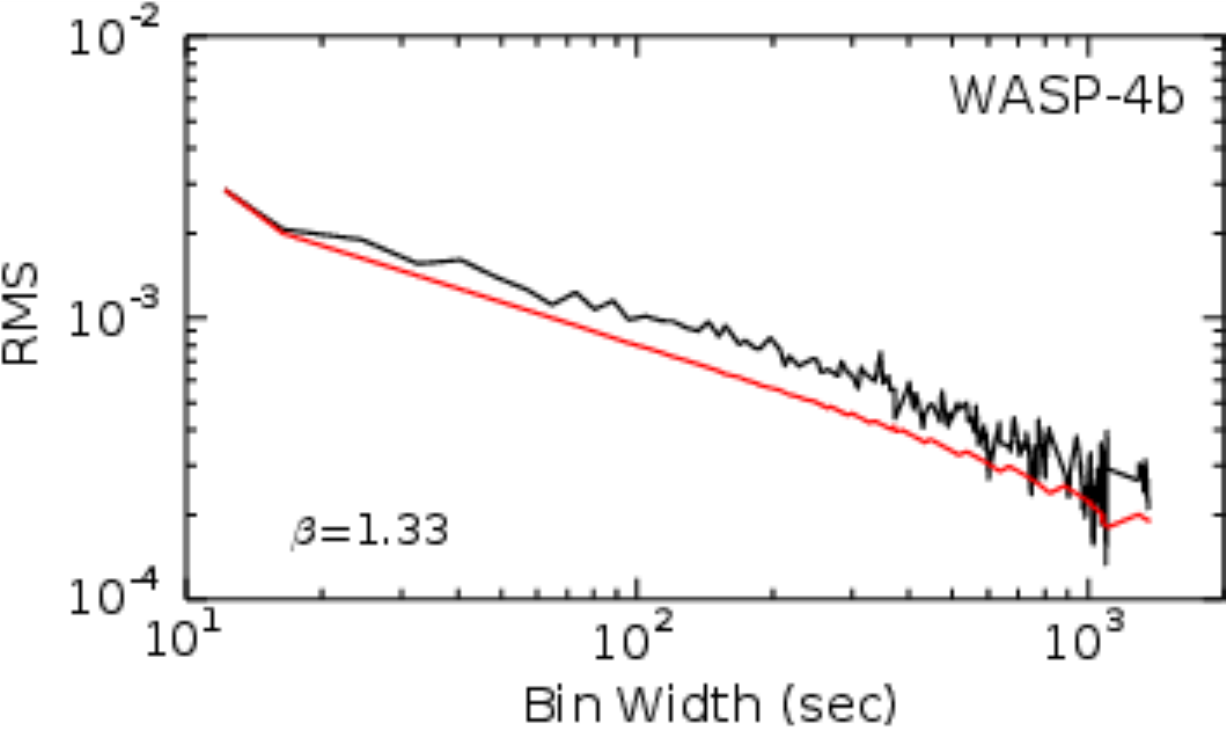} &
    \includegraphics[width=5cm]{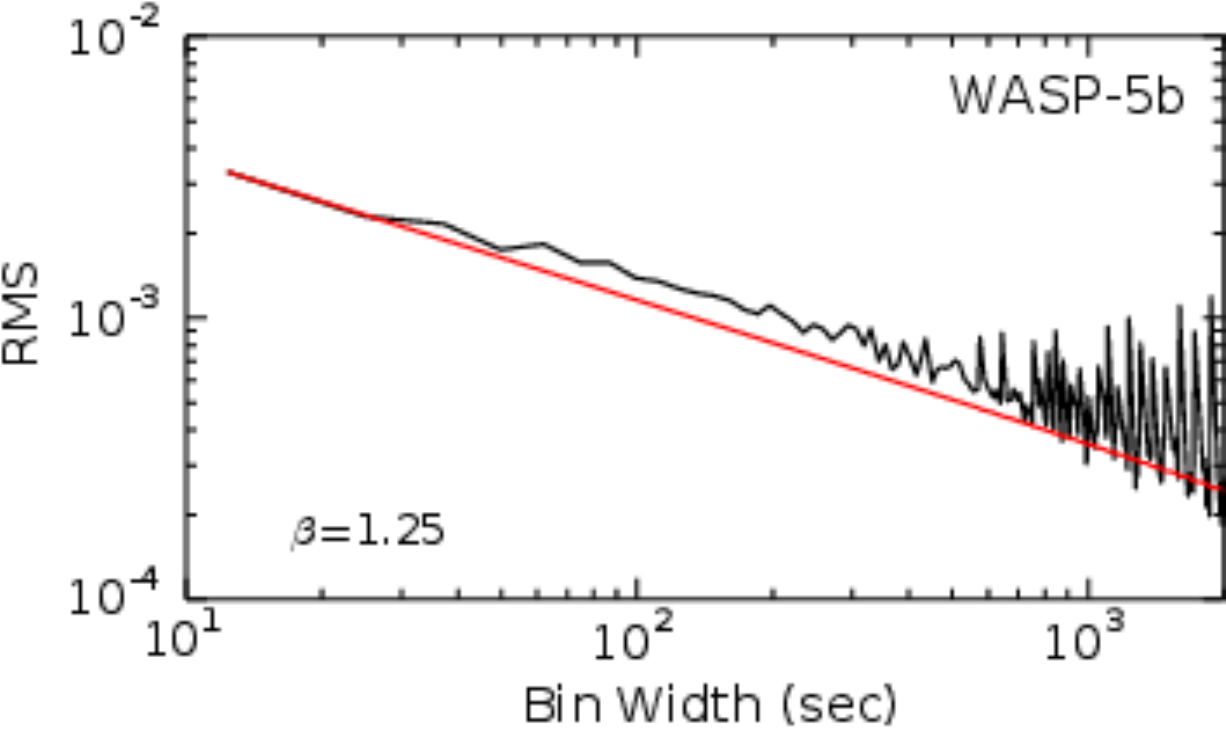} \\
    \includegraphics[width=5cm]{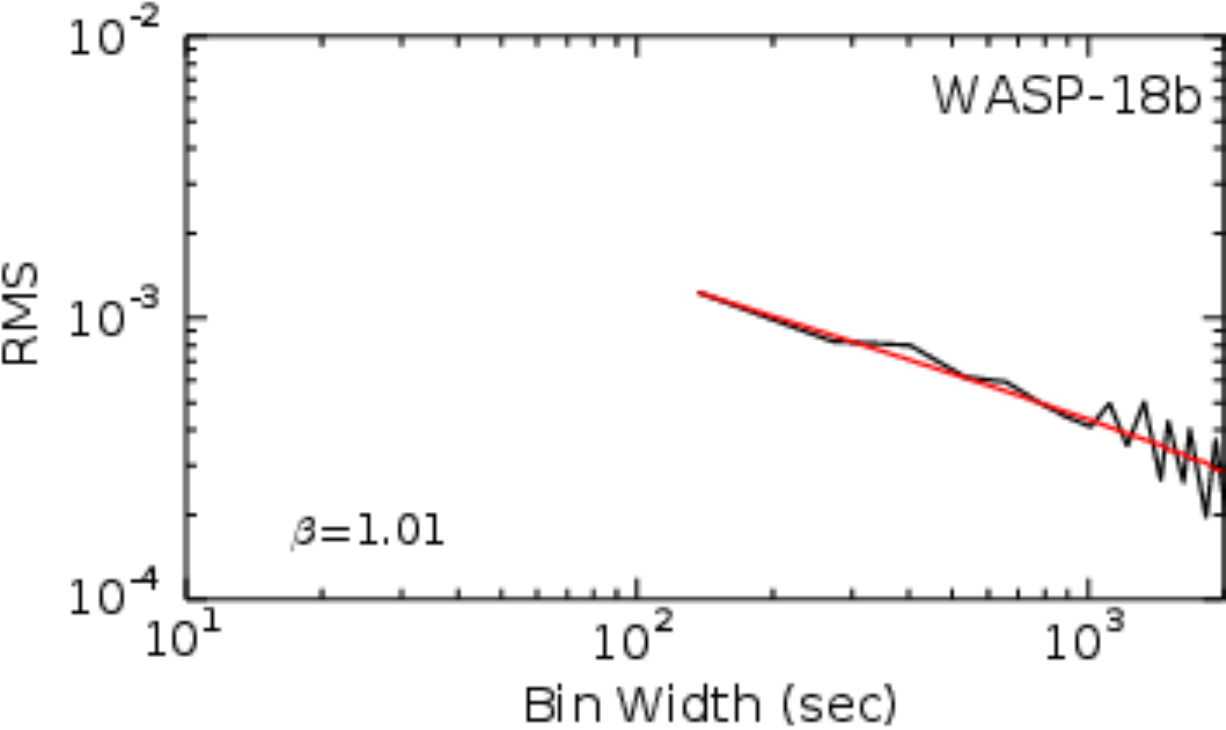} &
    \includegraphics[width=5cm]{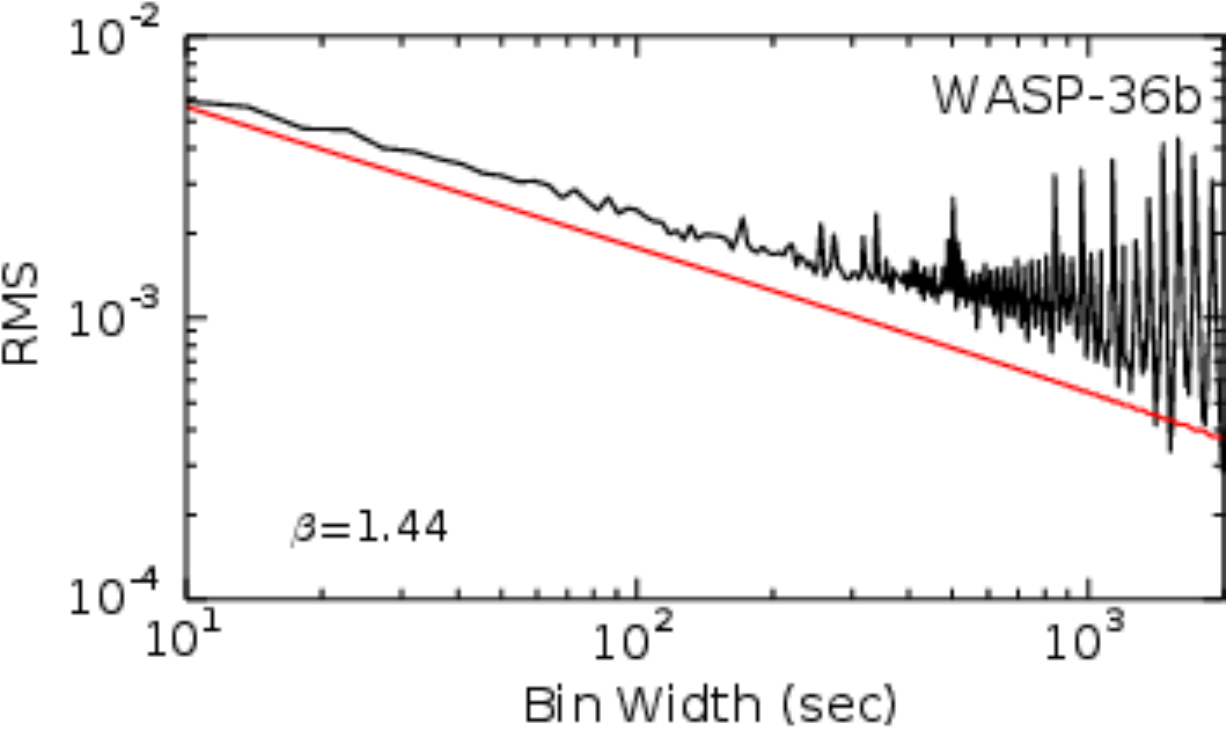} &
    \includegraphics[width=5cm]{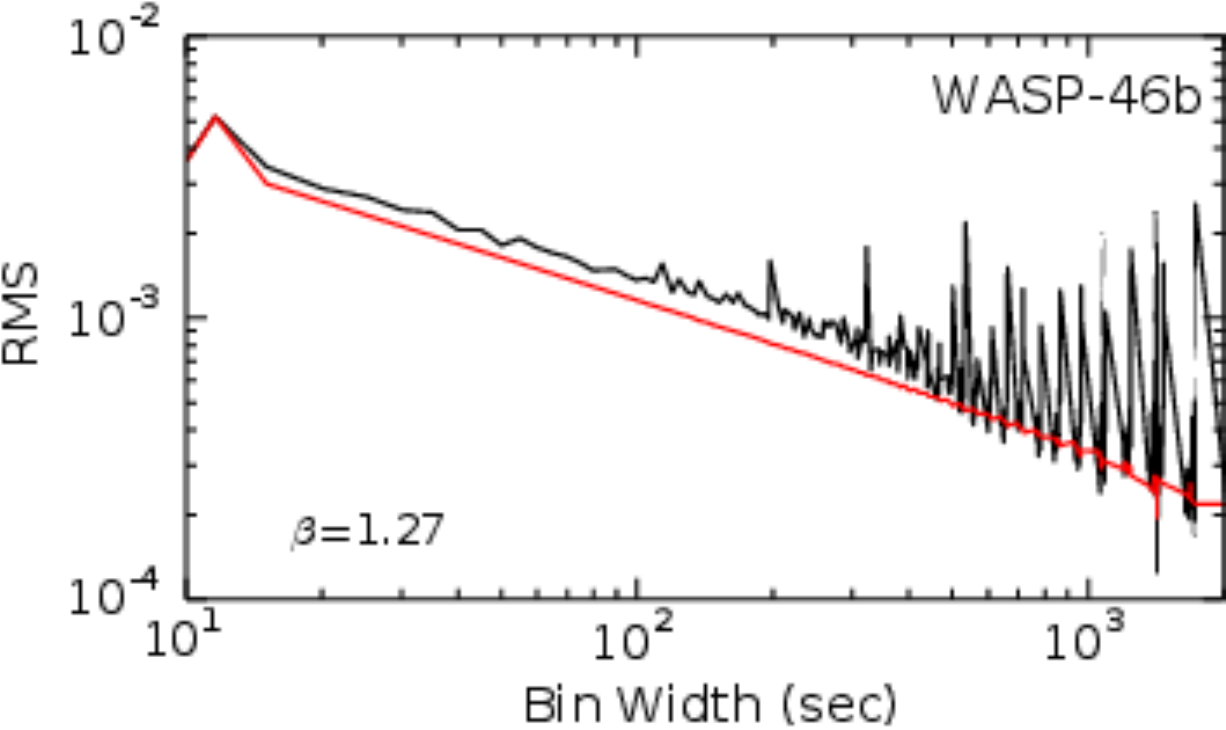} \\
    \includegraphics[width=5cm]{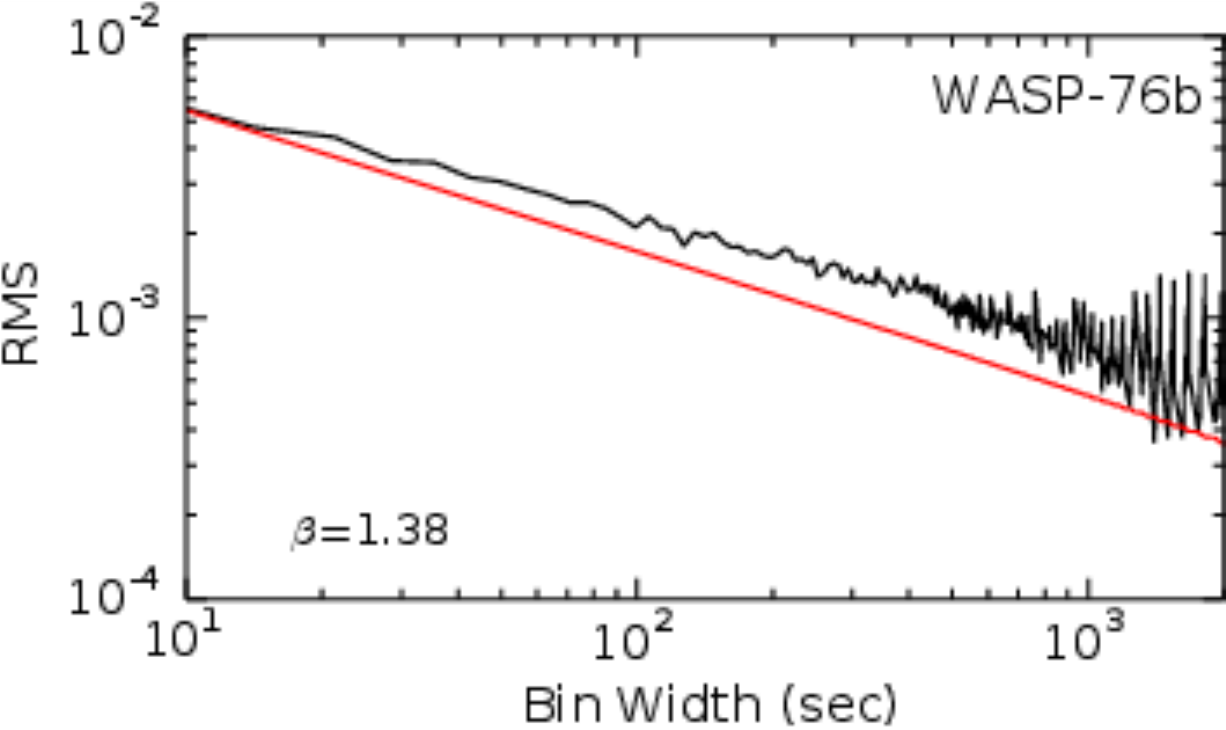} &&\\
  \end{tabular}
  \caption{The rms of the light curve residuals, with the eclipse and external parameter models subtracted, as a function of the bin sizes. The red line plots the expected rms assuming no time correlated noise, where the binned scatter scales $1/\sqrt{n}$. We calculate the average $\beta$ factor for each set of observations, which measures level of time correlation within the data binned at timescales between 60s and 600s. $\beta > 1$ indicates the presence of time-correlated noise in the residuals, while $\beta=1$ indicates residuals scale with the bin size.}
  \label{fig:lc_beta}
\end{figure*}

For each set of eclipses, we also calculate a $\beta$ factor to check for any residual time-correlated noise to the light curves \citep{2008ApJ...683.1076W}. We compared the progressive binned scatter of the light curve residual (i.e. the data with the eclipse and external parameter models subtracted) with the expected scatter assuming only photon noise. For every $m$ bins of $n$ points, we measure a root mean square (rms) scatter $\sigma_n$, which is then compared to the expected photon noise scaled rms from the unbinned light curve $(\sigma_1)$, according to:
\begin{equation}
  \label{eq:1}
  \sigma_n = \beta \frac{\sigma_1}{\sqrt{n}} \sqrt{\frac{m}{m-1}} \,.
\end{equation}
For light curves with no time correlated noise, $\beta=1$. The average $\beta$ value for each light curve, calculated for bins between 60s and 600s, along with the rms--bin size relationships, are shown in Figure~\ref{fig:lc_beta}. 

\begin{landscape}
\begin{table}
  \scriptsize
  \caption{Model fit and derived eclipse parameters}
  \label{tab:sys_param}

  \begin{tabular}{lrrrrrrrcrr}
    \hline\hline
    \multicolumn{6}{l}{\textbf{Model priors$^a$}} &  \multicolumn{3}{l}{\textbf{Model fit parameters }} &  \multicolumn{2}{l}{\textbf{Derived parameters}} \\
    Planet & \multicolumn{1}{l}{Period (days)} &\multicolumn{1}{l}{$T_0$ (BJD-TDB)$^b$} & \multicolumn{1}{l}{$(R_p+R_\star)/a$} & \multicolumn{1}{l}{$R_p/R_\star$} &  \multicolumn{1}{l}{$i\,(^\circ)$} & \multicolumn{1}{l}{$e\,\cos\omega$} & \multicolumn{1}{l}{$S_p/S_\star$}  & \multicolumn{1}{l}{Detrending} & \multicolumn{1}{l}{$F_p/F_\star$} & \multicolumn{1}{l}{$T_B$ (K)} \\
    & & & & &  & &  & \multicolumn{1}{l}{ Parameters$^c$} & Eclipse Depth (\%) & \\
    \hline
    WASP-2b & $2.1522214$ & $2453991.5153$ & $0.140\pm0.002$ & $0.1326\pm0.0007$ & $84.8\pm0.2$ & $-0.001_{-0.001}^{+0.001}$ $^d$ & $<0.042$ &  $A,t, Y$ & $<0.07$ & $<1900$\\
    & $\pm0.0000004$ & $\pm0.0002$ & &  & && & & & \\
    WASP-4b & 1.3382320 & 2454823.59192 & $0.211\pm0.001$ & $0.1545\pm0.0003$ & $88.5\pm0.4$ & $-0.001_{-0.003}^{+0.003}$ & $0.07_{-0.02}^{+0.02}$ & 2014-09-04: $t$ & $0.16_{-0.04}^{+0.04}$ & $1900_{-100}^{+100}$\\
    & $\pm0.0000002$ & $\pm0.00003$ & &  & && & 2014-09-11 $t,X$ & & \\
    WASP-5b & 1.628425 & 2454375.6257 & $0.203\pm0.007$ & $0.111\pm0.001$ & $86\pm1$ & $0.008_{-0.002}^{+0.002}$ & $0.16_{-0.02}^{+0.02}$ & $A,t$ & $0.20_{-0.02}^{+0.02}$ & $2500_{-100}^{+100}$ \\
    & $\pm0.000001$ & $\pm0.0002$ & &  & && & & & \\
    WASP-18b & $0.9414518$ & $2455084.79283$ & $0.306\pm0.009$ & $0.097\pm0.001$ & $85\pm2$ & $0.012_{-0.008}^{+0.007}$ & $0.14_{-0.03}^{+0.03}$ &  $t$ & $0.13_{-0.03}^{+0.03}$ & $2500_{-200}^{+200}$\\
    & $\pm0.0000004$ & $\pm0.00009$ & &  & && & & & \\
    WASP-36b & 1.537365 & 2455569.8381 & $0.190\pm0.003$ & $0.1384\pm0.007$ & $83.6\pm0.2$ & $0.004_{-0.005}^{+0.006}$ & $0.07_{-0.02}^{+0.02}$ & $A,t,Y$ & $0.13_{-0.04}^{+0.04}$ & $1900_{-200}^{+100}$\\
    & $\pm0.000003$ & $\pm0.0001$ & &  & && & & & \\
    WASP-46b & 1.430370 & 2455392.3163 & $0.200\pm0.006$ & $0.147\pm0.007$ & $82.6\pm0.4$ & $0.004_{-0.002}^{+0.002}$ & $0.12_{-0.02}^{+0.02}$ & 2014-09-11: $A,t$ & $0.26_{-0.03}^{+0.05}$ & $2200_{-100}^{+100}$\\
    & $\pm0.000002$ & $\pm0.0002$ & &  & && & 2014-09-14: $A,t,F$ & & \\
    WASP-76b & 1.809886 & 2456107.8551 & $0.270\pm0.007$ & $0.1090\pm0.0007$ & $88\pm2$ & $-0.001_{-0.004}^{+0.004}$ $^e$ & $<0.26$ & $A,F,t,Y$ & $<0.3$ & $<3500$\\
    & $\pm0.000001$ & $\pm0.0003$ & &  & && & & & \\

    \hline
  \end{tabular}
\begin{flushleft} 
$^a$ Priors adopted from literature system parameters: WASP-2b \citep{2007MNRAS.375..951C,2010MNRAS.408.1680S}, WASP-4b\citep{2008ApJ...675L.113W,2013MNRAS.434...46H}, WASP-5b \citep{2008MNRAS.387L...4A,2009MNRAS.396.1023S}, WASP-18b \citep{2009Natur.460.1098H,2009ApJ...707..167S}, WASP-36b \citep{2012AJ....143...81S}, WASP-46b \citep{2012MNRAS.422.1988A}, WASP-76b \citep{2013arXiv1310.5607W}\\
$^b$ Where approporiate, transit times reported in HJD-UTC have been translated to BJD-TDB using \citet{2010PASP..122..935E}.\\
$^c$ The set of detrending parameters are airmass $A$, background flux $B$, FWHM $F$, linear time dependent trend $t$, target X pixel positions $X$ and $Y$.\\
$^d$ $e\cos\omega$ for WASP-2b constrained by Gaussian prior from Spitzer eclipse measurements to be $-0.0013\pm0.0009$ \citep{2010arXiv1004.0836W}.\\
$^e$ $e\cos\omega$ for WASP-76b constrained by Gaussian prior from radial velocity measurements to be $0.00\pm0.02$ \citep{2013arXiv1310.5607W}.\\
\end{flushleft}
\end{table}
\end{landscape}


\section{Discussion}
\label{sec:discussion}

We report new $Ks$ band eclipse depth measurements and constraints for seven hot-Jupiters. These results bring the total number of planets with eclipses monitored at $2.1\,\mu\text{m}$ to 25, a sample large enough to allow some initial statistical insight into the atmospheres of the hot-Jupiter population. 

Such statistical analyses require a set of robust measurements with reliable uncertainty estimates. To characterise the robustness and repeatabilitiy of our measurements, we performed an independent analysis for the two full eclipses of WASP-46b that we obtained (Figure~\ref{fig:eclipse_lc_WASP46}). We derived self-consistent eclipse phases and depths between the two eclipses to within $1\sigma$ ($e\cos\omega$ of $0.006_{-0.005}^{+0.005}$ and $0.001_{-0.004}^{+0.004}$, and eclipse depth $F_p/F_\star$ of $0.25_{-0.05}^{+0.05}$\% and $0.32_{-0.05}^{+0.06}$\%). On both nights, the eclipses were detected at $>5\sigma$ significance, and the measured eclipse parameters were consistent with each other at the $1\sigma$ level. Literature $Ks$ band eclipse measurements also exist for WASP-4b, -5b, and -46b. The eclipse depths we report are also consistent within $1\sigma$ to all the previous measurements. This increases our confidence of the uncertaintiy measurements presented by the series of ground-based eclipse observations to date. For the wider sample of literature eclipse measurements, the uncertainties in the eclipse depths are also often underestimated: the scatter in repeated eclipse depths reported for the same planet is 1.4 times larger than the mean error estimates for $Ks$-band observations \citep{2014MNRAS.445.2746Z}, and two times larger for \emph{Spitzer} measurements \citep{2014MNRAS.444.3632H}.

However, eclipse measurements remain intrinsically difficult due to the low signal-to-noise nature of the planetary eclipse and the variety of systematic signals that can be introduced. Time-correlated noise often still remains in our observations despite decorrelation against the instrumental model, as most of the light curve residuals have $\beta > 1$. In cases where the time-correlated noise is significant (e.g. observations of WASP-36b and WASP-76b), we notice the uncertainties in the eclipse depth are larger than other eclipses of equivalent depths. 

With these cavaets, we can place our observations in the context of the hot-Jupiter atmospheres sample. In this section, we empirically examine the colour-magnitude distributions and brightness-equilibrium temperature distributions of the hot-Jupiter population.

\subsection{Colour-magnitude diagram}
\label{sec:colo-magn-diagr}
\begin{figure*}
  \centering
  \includegraphics[width=16cm]{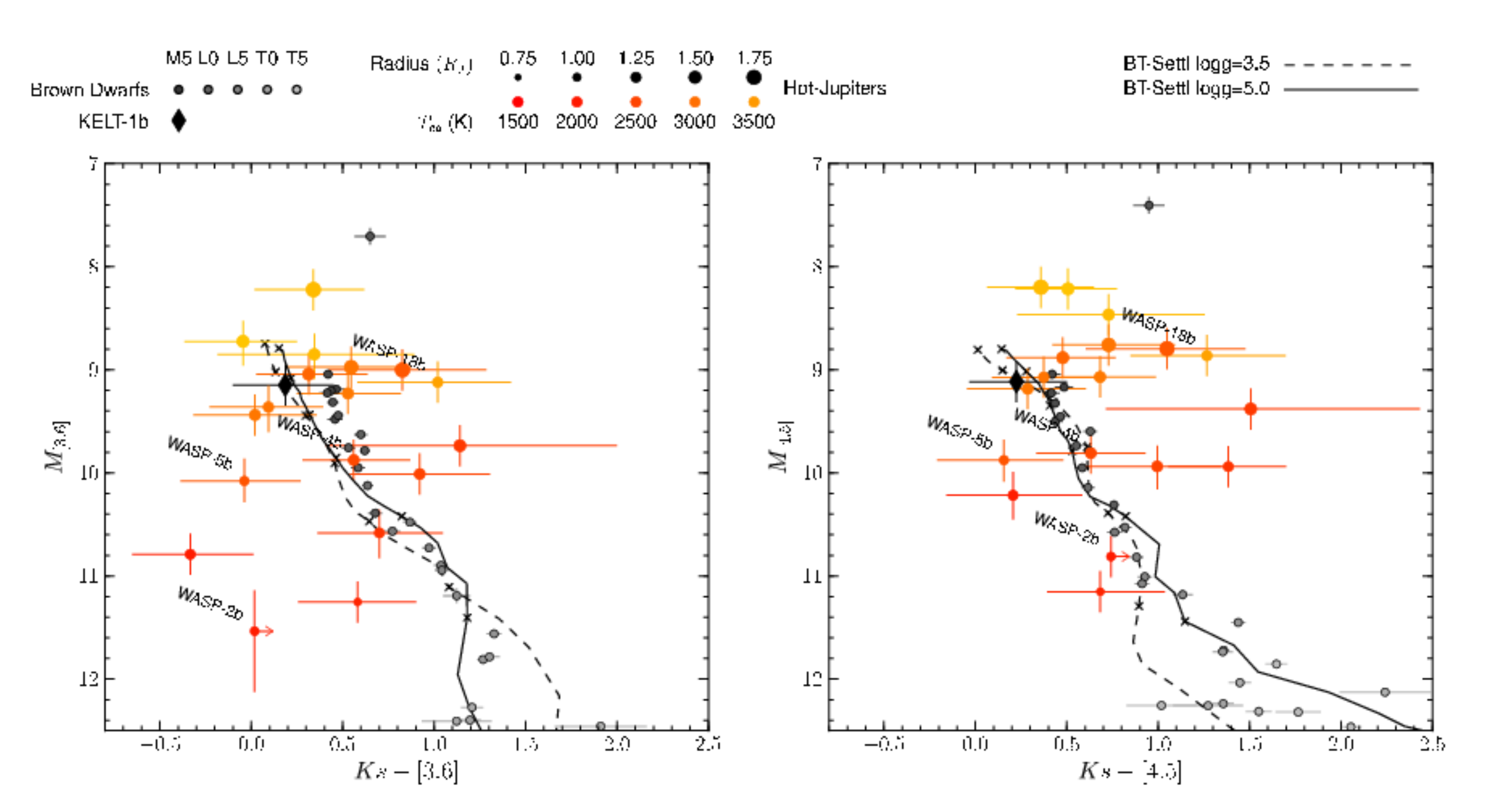}
  \caption{Colour-magnitude diagrams for hot-Jupiters and brown dwarfs. Hot-Jupiters are plotted in colour according to their equilibrium temperature, and with point sizes relative to their planet radii. The planets with $Ks$ band eclipses reported in this paper, and with available \emph{Spitzer} eclipse observations, are labelled. Where appropriate, upper limits in the colour axis are given. The transiting brown dwarf KELT-1b is plotted as the black diamond. It is the only irradiated brown dwarf with secondary eclipse measurements available. The colour-magnitudes of brown dwarfs with parallaxes and photometry compiled by \citet{2012ApJS..201...19D} are plotted for comparison, the points are gray-scaled according to spectral class. The BT-Settl model colours, from $<4000$\,K, are marked by the black lines. The corresponding effective temperatures of the models are marked by crosses every 500\,K. Models at $\log g=3.5$ and 5.0, corresponding to the surface gravities of Jupiter and typical brown dwarfs, are marked by the dashed and solid lines respectively. The absolute magnitudes for the models are converted assuming $1\,R_J$ objects. References for the parallaxes of the points plotted are \citet{2011AJ....141...54A,2010ApJ...718L..38A,1999AJ....118.1086B,2005AJ....130..337C,2006AJ....132.1234C,2002AJ....124.1170D,2012ApJS..201...19D,2009AJ....137..402G,1988AJ.....95.1841G,1993AJ....105.1571H,2006AJ....132.2360H,2011ApJS..197...19K,2009AJ....137.4109L,2010A&A...524A..38M,1992AJ....103..638M,2003AJ....125..354R,2009A&A...493L..27S,2009AJ....137.4547S,2008A&A...489..825T,1995AJ....110.3014T,1996MNRAS.281..644T,2003AJ....126..975T,1995gcts.book.....V,2007ASSL..350.....V,2004AJ....127.2948V}. 
References for the 2MASS photometry used are \citet{2010ApJ...720L..82B,2011A&A...528L..15B,2006AJ....131.2722C,2002ApJ...567L..53C,2011ApJ...736L..33C,2003tmc..book.....C,2012ApJ...755...94D,2009ApJ...706..328D,2009ApJ...699..168D,2010ApJ...721.1725D,2012ApJS..201...19D,2013A&A...549A..52E,1988ApJ...330L.119F,2011ApJ...739L..41G,2004AJ....127.3516G,1993AJ....106..773H,2006MNRAS.367..454H,2011ApJ...728...85J,1996MNRAS.280...77J,2007A&A...471..655K,2010A&A...510A..99K,2011ApJS..197...19K,2004AJ....127.3553K,2010ApJ...711.1087K,2001ApJ...560..390L,1998ApJ...509..836L,2000ApJ...536L..35L,2001ApJ...548..908L,2002ApJ...564..452L,2002MNRAS.332...78L,2007ApJ...655.1079L,2006ApJ...647.1393L,2008ApJ...689..436L,2010ApJ...722..311L,2007MNRAS.379.1423L,2000ApJ...541..390L,2008Sci...322.1348M,2010Natur.468.1080M,2006ApJ...651.1166M,2007MNRAS.378.1328M,2012ApJ...750...53N,2002AJ....123.2806R}.
References for the IRAC photometry used are \citet{2011ApJS..197...19K,2007ApJ...655.1079L,2010ApJ...722..311L,2012ApJ...744..135L,2006ApJ...651..502P,2010AJ....140.1868W}. }
  \label{fig:colormag}
\end{figure*}

We compare the broadband colours and magnitudes of the planets examined in this study with other hot-Jupiters, brown dwarfs, and late M-dwarfs. Brown dwarfs and late M-dwarfs have similar effective temperatures as the equilibrium temperatures of hot-Jupiters, but higher surface gravities, and experience different levels of irradiation to hot-Jupiters. In addition, brown dwarf atmospheres are relatively better understood, and brown dwarf spectral models are the source from which most hot-Jupiter atmosphere models are built. Colour-magnitude and colour-colour diagrams can help compare the broadband spectra hot-Jupiters, and examine for differences between the atmospheres of hot-Jupiters and brown dwarfs \citep[e.g.][]{2008Sci...322.1348M,2014MNRAS.439L..61T,2014MNRAS.444..711T}. 

To calculate the absolute magnitudes of the hot-Jupiters, we first derive absolute magnitudes for their host stars. The published effective temperature, surface gravity, and metallicity of the host stars are fitted to the Dartmouth isochrones \citep{2008ApJS..178...89D}, from which the absolute magnitudes at $Ks$ and the \emph{Spitzer} IRAC bands are extracted. To propagate the uncertainties, we draw $10^3$ iterations of the stellar atmospheric parameters from Gaussian distributions about their reported mean and uncertainty values. For the seven host stars that have Hipparcos distances \citep[identified in ][]{2014MNRAS.439L..61T}, the isochrone derived $J$, $H$, $K$ distance modulus agree with that measured from parallax, with 0.2 mag scatter in the residuals. We set 0.2 mag as the lower limit for the absolute magnitude uncertainties we calculate from isochrones. We then calculate the absolute magnitudes of the planets using the measured eclipse depth values and uncertainties. In the cases where repeated observations are available, we calculate a weighted mean eclipse depth and a  weighted standard error in the mean for the uncertainty. The literature eclipse measurements are listed in Appendix~\ref{sec:liter-eclipse-meas}, and gathered partially from the Exoplanet Orbit Database\footnote{This research has made use of the Exoplanet Orbit Database and the Exoplanet Data Explorer at exoplanets.org.} \citep{2014PASP..126..827H} and Table 3 from \citet{2014PASA...31...43B}. Hot-Jupiters in high eccentricity orbits (HD 80606b, WASP-8b) were excluded from the list as they are not representative of the hot-Jupiter sample. 

Figure~\ref{fig:colormag}  plots the $Ks-[3.6]$ and $Ks-[4.5]$ colours of the hot-Jupiters and brown dwarfs against the $M_{[3.6]}$ and $M_{[4.5]}$ band absolute magnitudes. These bands are chosen as they have the most number of eclipse measurements. The hot-Jupiters are plotted in colour to represent their equilibrium temperatures, with point sizes indicating their planet radii. Brown dwarfs compiled from \citet{2012ApJS..201...19D}  are plotted in gray scale to represent their spectral classes. Model colours from BT-Settl \citep{2012RSPTA.370.2765A}, with abundances from \citet{2009ARA&amp;A..47..481A}, are plotted for reference. 

The sampled hot-Jupiters reside around the M--L spectral classes. The colours of hot-Jupiters are consistent with the colours of brown dwarfs. The  $Ks-[4.5]$ colours are marginally redder for the hot-Jupiters than brown dwarfs. A discrepancy between the two populations based on $4.5\,\mu\text{m}$-related colours was suggested by \citet{2014MNRAS.444..711T}, with the mechanism being the absence of absorbing spectral features at $4.5\,\mu\text{m}$ for hot-Jupiters compared to brown dwarfs. The transiting irradiated brown dwarf KELT-1b \citep{2012ApJ...761..123S} is also plotted. It is the only brown dwarf, receiving similar irradiation levels as hot-Jupiters, with secondar eclipse measurements. The colours of KELT-1b matches well with that of isolated brown dwarfs and brown dwarf atmosphere models.

\begin{figure*}
  \centering
  \includegraphics[width=13cm]{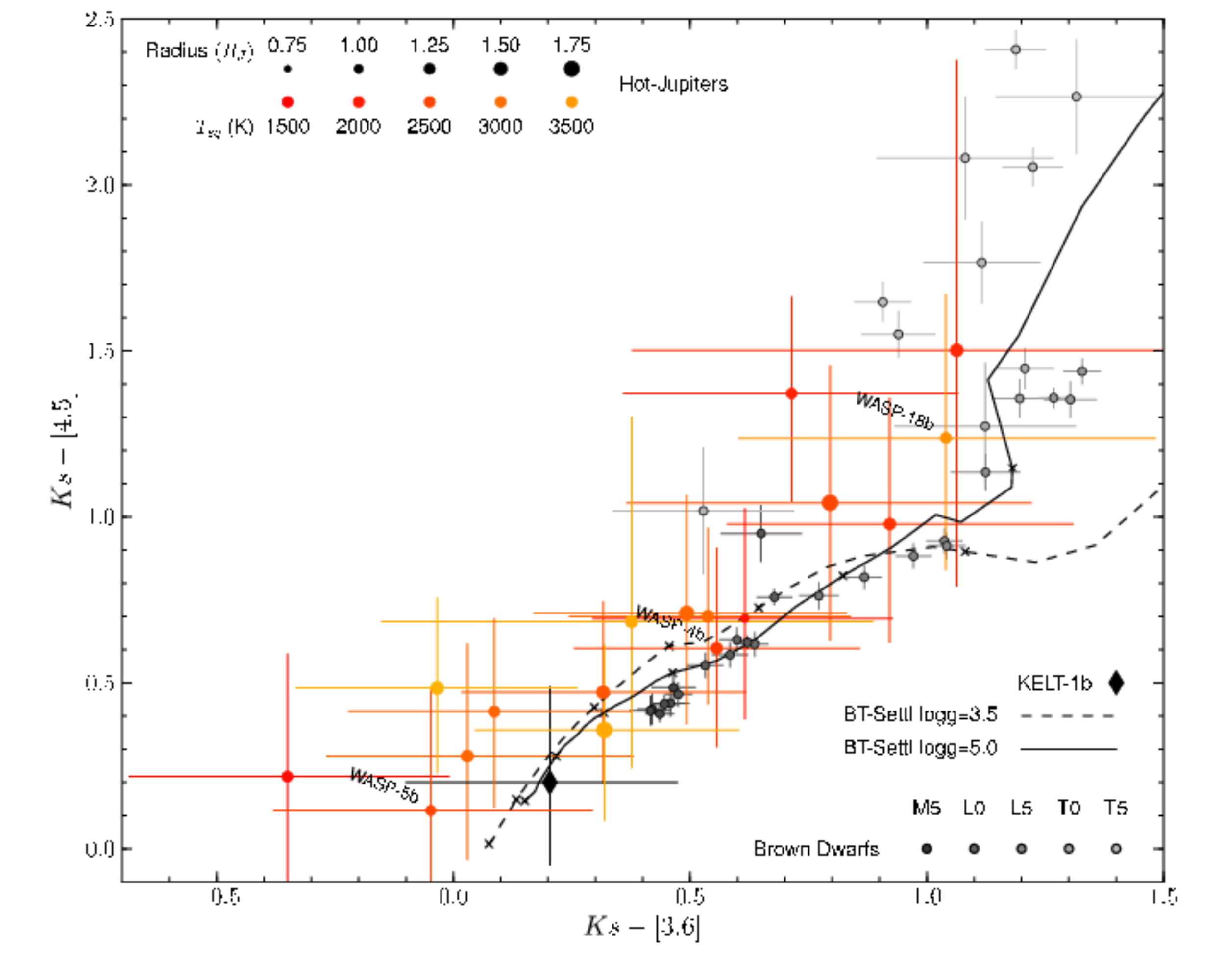}
  \caption{$Ks-[3.6]$ vs $Ks-[4.5]$ colour-colour diagram for hot-Jupiters and brown dwarfs. Colour-colour diagrams reduce the effect of the large scatter in planet radii, therefore luminosity, on the spectral class comparisons. The plot markings and references are the same as Figure~\ref{fig:colormag}. }
  \label{fig:colorcolor}
\end{figure*}

For a direct comparison of the spectral properties of hot-Jupiters and brown dwarfs, we plot their $Ks-[3.6]$ vs $Ks-[4.5]$ colour-colour relationship in Figure~\ref{fig:colorcolor}. This removes the luminosity dependence on radius and reduces the scatter in the distribution. Here, it is interesting to note that the current sample of hot-Jupiter colours matches well with that of brown dwarfs.

We expect a greater diversity in the hot-Jupiter spectral properties compared to brown dwarfs. Hot-Jupiters have radii that differ by a factor of two between similar mass planets, and have atmospheres that are heated from above and below at different levels of irradiation, resulting in a wide range of possible pressure-temperature profiles. We therefore expect the colour distribution of hot-Jupiters to exhibit significantly greater scatter than that of brown dwarfs. However, the scatter in the colour distribution is currently dominated by the measurement uncertainties of the eclipse observations. Further repeated observations, for robust colour measurements, and a greater sample size, may help distinguish between the brown dwarf and planet population.

\subsection{Brightness -- equilibrium temperature distribution}
\label{sec:brightn-equil-temp}

\begin{figure*}
  \centering
  \includegraphics[width=16cm]{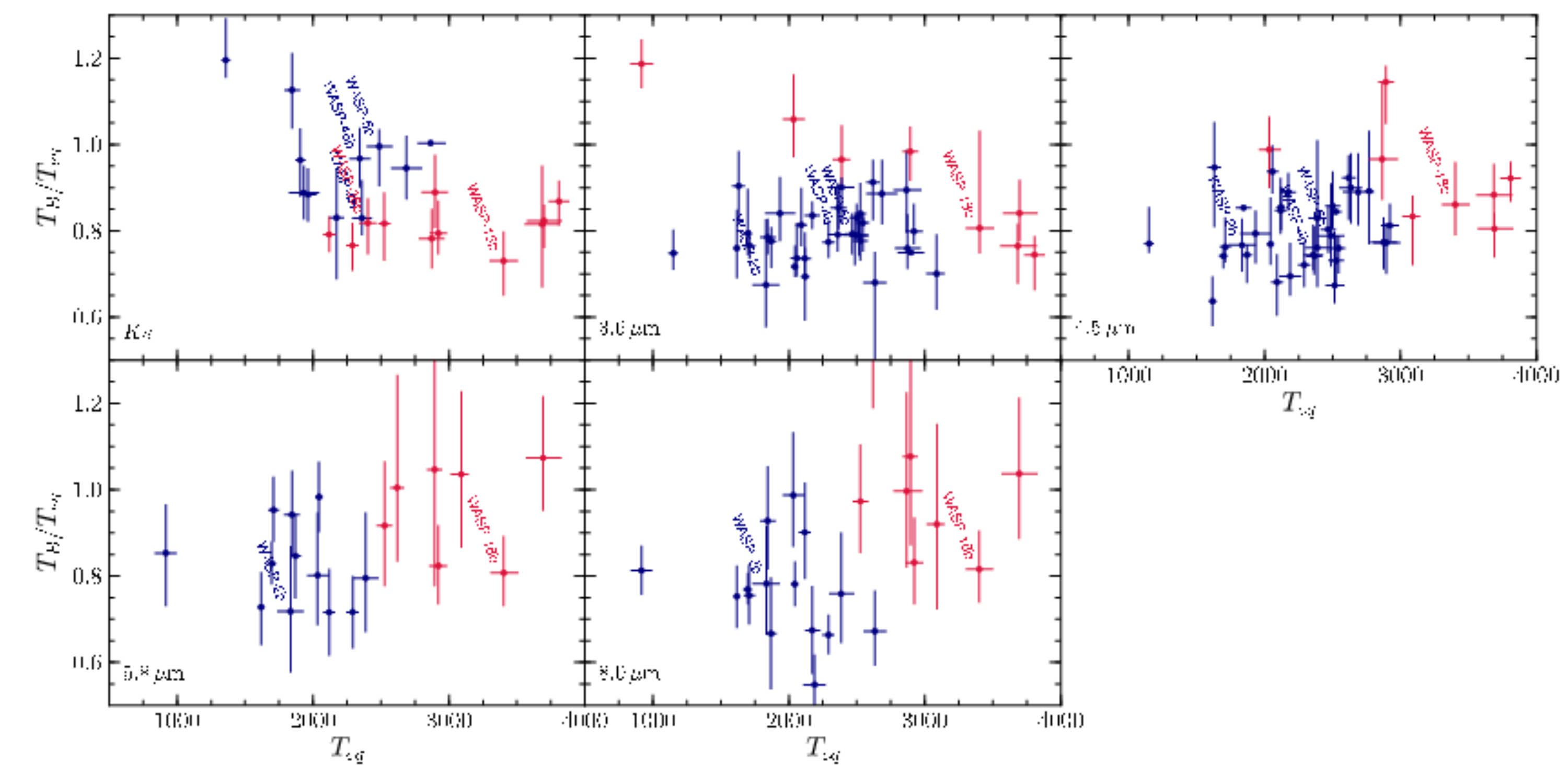}
  \caption{The normalised brightness temperature $T_B/T_{eq}$ -- equilibrium temperature $T_{eq}$ distribution from secondary eclipses measured at near infrared bands. The hot-Jupiters with $Ks$ band eclipses reported in this paper, and with relevant \emph{Spitzer} band observations, are labelled. Gaussian Mixture Model (GMM) clustering preferentially selects a single component model for the distribution at all the bands. The blue and red colours show the clusters if we force a $N=2$ component fit to the distributions.}
  \label{fig:teq_tb}
\end{figure*}

\begin{figure*}
  \centering
  \includegraphics[width=16cm]{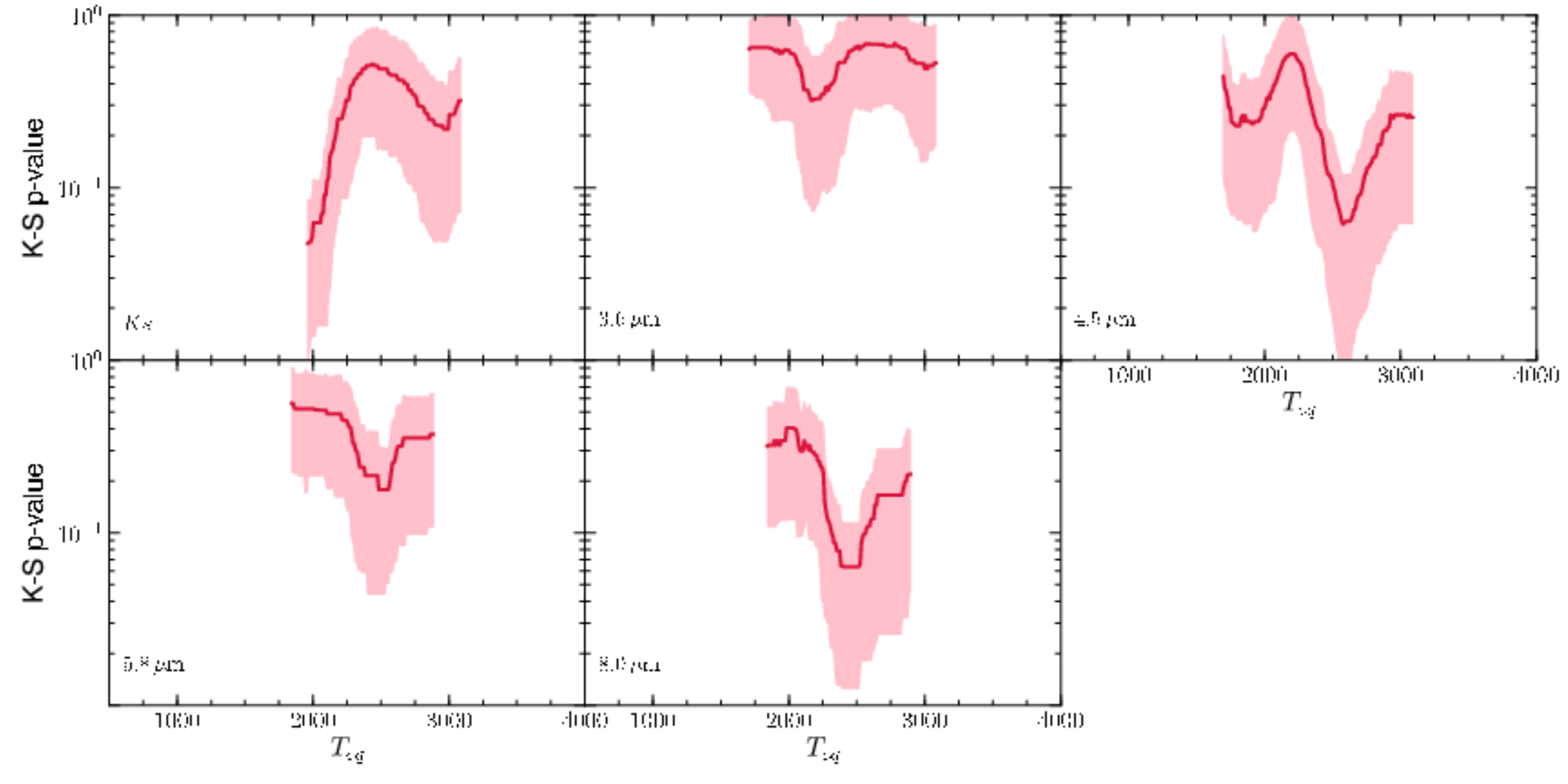}
  \caption{The p-value -- $T_{eq}$ relation from a moving K-S test along the $T_B/T_{eq}$ -- $T_{eq}$ distribution. Minima in the p-values should indicate divisions between distinct populations in the sample. We find no significant p-value minima (all $>0.05$), indicating a lack of significant division in the population at all the bands. }
  \label{fig:ks_pvalue}
\end{figure*}

3D models investigating the circulation of strongly irradiated hot-Jupiters have predicted large day-night temperature differences, with the most irradiated planets developing strong, super-rotating, equatorial jets and large longitudinal temperature gradients \citep[e.g.][]{2002A&amp;A...385..166S,2008ApJ...673..513D}. These have been revealed by observations of infrared phase curves with peaks offset from the sub-stellar point \citep[e.g.][]{2007Natur.447..183K,2014ApJ...790...53Z,2014Sci...346..838S}. \citet{2015ApJ...801...95S} have shown the extent of this day-night temperature difference is dependent on the level of irradiation and the rotation rate of the planet. In comparison, the circulation of mildy irradiated and/or rapidly rotating hot-Jupiters are expected to be dominated by latitudinal variation, and weaker longitudinal differences. 

To probe for a boundary between the mildy irradiated, thermally well mixed hot-Jupiters, and the strongly irradiated hot-Jupiters, \citet{2011ApJ...729...54C} and \citet{2015arXiv150206970S} used the available multi-band eclipse observations to compare day-side effective temperatures of hot-Jupiters with their expected equilibrium temperatures. They tentatively identified two populations of hot-Jupiters, with the most irradiated planets having lower heat recirculation efficiencies, and higher relative effective temperature, than mildly irradiated hot-Jupiters.

We re-examine this proposed dichotomy for the photometric bands where a significant number of hot-Jupiters have been sampled in eclipse. Each infrared band probes a different layer of the planetary atmosphere, with shorter wavelengths probing higher pressure regions, where models predict better thermal mixing than at higher altitudes. For each planet sampled in each of the $Ks$ and \emph{Spitzer} IRAC bands, we calculate an equilibrium temperature $T_{eq}$, assuming zero albedo and no heat redistribution. The uncertainty on the equilibrium temperature is calculated from $10^3$ iterations of random sampling, such that errors in the semi-major axis, stellar radius, and stellar effective temperature are propagated. For each band, we calculate a brightness temperature $T_B$ from the reported average eclipse depths (listed in  Appendix~\ref{sec:liter-eclipse-meas}). The $T_{eq}$ of each planet and each band is plotted against the normalised $T_B/T_{eq}$ ratio in Figure~\ref{fig:teq_tb}. 

To test for multiple populations within the $T_B/T_{eq}$ distribution, we apply the moving two-sample Kolmogorov-Smirnov (K-S) test and the Gaussian Mixture Model (GMM) analyses. These techniques have previously been used to check for distinct populations in the metallicity -- planet radius \citep{2014Natur.509..593B,2015ApJ...799L..26S} and metallicity -- planet period distributions \citep{2015arXiv150301771Z}. 

In a moving K-S test, we split the population into two samples along a series of $T_{eq}$ values, and in each instance calculate the K-S test p-values to test the null hypothesis that the $T_B/T_{eq}$ distribution for the two samples originate from the same population. If a significant minimum is observed in the $T_{eq}$ -- p-value relationship, then we can state that the distribution is made of two distinct populations. To propagate the uncertainties in $T_{eq}$ and $T_B$, we perform the moving K-S test $10^3$ times, at each iteration drawing each point from distributions about its mean and error. The $T_{eq}$ -- p-value relationship from the moving K-S test is plotted in Figure~\ref{fig:ks_pvalue}. We recover the weak division suggested by \citet{2011ApJ...729...54C} in all the \emph{Spitzer} bands, finding a tentative division at 2170, 2590, 2440, and 2380\,K in the 3.6, 4.5, 5.8, and $8.0\,\mu\text{m}$ bands. There appears to be no significant division in the $Ks$ band, with the lack of a clear minimum in the $T_{eq}$ -- p-value relationship. However, the division is statistically insignificant in all of the bands, with the minimum p-value consistently $\geq 0.05$. 

\citet{2015ApJ...799L..26S} suggested modelling the populations as Gaussian mixtures as a more rigorous way of distinguishing between multiple clusters in the population. We employ the GMM clustering function in the Python package Scikit-learn \citep{2012arXiv1201.0490P}. We fit the $T_{eq}$--$T_B/T_{eq}$ distribution with GMMs consisting of $N=1,2...,5$ full Gaussian components. The model that minimises the BIC is chosen as the best fit model. To take into account the per point uncertainties, we draw the population from their measurement uncertainties $10^3$ times, each time performing the BIC calculation and model selection. We find that the single component model is preferred for the $T_{eq}$--$T_B/T_{eq}$ distribution at every band. The $N\geq 2$ component models are rejected $>90$\% of the time in the $Ks$, 4.5, 5.8, and $8.0\,\mu\text{m}$ bands, and $>60$\% of the time for the $3.6\,\mu\text{m}$ band. We find a lack of conclusive evidence that the eclipse sample can be split into two populations. 

Given the limits of the current data, these tests show that the atmospheric circulation properties of the hot-Jupiter population is continuous. We suggest that there is likely no sharp divide between the warm Jupiters dominated by latitudinal circulation and the hot-Jupiters with longitudinal circulation.

However, if we were to force a two component fit to the GMM, we find a relatively consistent result between the groups identified in each band. There is also a consistency in the 4.5, 5.8, and $8.0\,\mu\text{m}$ bands between the divisions identified by the moving K-S test and the division between clusters found by GMM. Figure~\ref{fig:teq_tb} colour-codes each planet according to the classifications from the two-component GMM fit. 

The difference between the mean $T_B/T_{eq}$ between the two clusters is increasing with wavelength. The difference in the mean is $0.03\pm0.18$ at the $Ks$ band, and $0.24\pm0.20$ at [8.0]. Longer wavelength probe the upper planetary atmosphere, where the day-night temperature gradient is expected to be highest, whilst better thermal mixing is expected at higher pressures deeper in the atmosphere. The same effect should also lead to a larger difference between the two proposed hot-Jupiter populations at longer wavelengths, as demonstrated by this trend. We suspect that an underlying smooth transition may exist, but must await a larger sample size before becoming statistically significant.

\section*{Acknowledgements}
\label{sec:acknowledgements}
We thank the support of the AAO staff who helped with establishing the observing strategy employed in this work. CGT gratefully acknowledges the support of ARC Australian Professorial Fellowship grant DP0774000 and ARC Discovery Outstanding Researcher Award DP130102695.

\appendix

\section{Literature eclipse measurements}
\label{sec:liter-eclipse-meas}

Table~\ref{tab:lit_param} presents literature eclipse depth measurements for the $Ks$, \emph{Spitzer} IRAC 3.6, 4.5, 5.8, $8.0\,\mu\text{m}$ bands. 



\onecolumn
{\footnotesize \begin{longtable}{lp{2cm}p{2cm}p{2cm}p{2cm}p{2cm}p{4cm}}

\caption{Literature secondary eclipses in the $Ks$ and \emph{Spitzer} IRAC bands } \label{tab:lit_param}\\

\hline\hline Planet & \multicolumn{1}{l}{$Ks$ depth (\%)} &\multicolumn{1}{l}{[3.6] depth (\%)} & \multicolumn{1}{l}{[4.5] depth (\%)} & \multicolumn{1}{l}{[5.8] depth (\%)} & \multicolumn{1}{l}{[8.0] depth (\%)} & References \\ \hline 
\endfirsthead

\multicolumn{7}{c}%
{ \tablename\ \thetable{} -- continued from previous page} \\
\hline\hline Planet & \multicolumn{1}{l}{$Ks$ depth (\%)} &\multicolumn{1}{l}{[3.6] depth (\%)} & \multicolumn{1}{l}{[4.5] depth (\%)} & \multicolumn{1}{l}{[5.8] depth (\%)} & \multicolumn{1}{l}{[8.0] depth (\%)} & References \\ \hline 
\endhead

\hline \multicolumn{7}{r}{{Continued on next page}} \\ \hline
\endfoot

\hline
\endlastfoot

55 Cnc e &  &  & $0.0131_{-0.0028}^{+0.0028}$  &  &  & {\citet{2012ApJ...751L..28D}}\\ 
CoRoT-1b & $0.278_{-0.066}^{+0.043}$ $0.336_{-0.042}^{+0.042}$  & $0.415_{-0.042}^{+0.042}$  & $0.482_{-0.042}^{+0.042}$  &  &  & {\citet{2009A&amp;A...506..359G,2009ApJ...707.1707R,2011ApJ...726...95D}}\\ 
CoRoT-2b & $0.16_{-0.09}^{+0.09}$  & $0.355_{-0.02}^{+0.02}$  & $0.51_{-0.042}^{+0.042}$ $0.5_{-0.02}^{+0.02}$  &  & $0.41_{-0.11}^{+0.11}$ $0.446_{-0.1}^{+0.1}$  & {\citet{2010AJ....139.1481A,2011ApJ...726...95D,2010A&amp;A...511A...3G}}\\ 
GJ436b &  & $0.041_{-0.003}^{+0.003}$  & $<0.01$  & $0.033_{-0.014}^{+0.014}$  & $0.057_{-0.008}^{+0.008}$ $0.054_{-0.007}^{+0.007}$ $0.054_{-0.008}^{+0.008}$ $0.0452_{-0.0027}^{+0.0027}$  & {\citet{2010Natur.464.1161S,2007ApJ...667L.199D,2007A&amp;A...475.1125D,2011ApJ...735...27K}}\\ 
HAT-P-1b & $0.109_{-0.025}^{+0.025}$  & $0.08_{-0.008}^{+0.008}$  & $0.135_{-0.022}^{+0.022}$  & $0.203_{-0.031}^{+0.031}$  & $0.238_{-0.04}^{+0.04}$  & {\citet{2011A&amp;A...528A..49D,2010ApJ...708..498T}}\\ 
HAT-P-2b &  & $0.0996_{-0.0072}^{+0.0072}$  & $0.1031_{-0.0061}^{+0.0061}$  & $0.071_{-0.013}^{+0.029}$  & $0.1392_{-0.0095}^{+0.0095}$  & {\citet{2013ApJ...766...95L}}\\ 
HAT-P-3b &  & $0.112_{-0.03}^{+0.015}$  & $0.094_{-0.009}^{+0.094}$  &  &  & {\citet{2013ApJ...770..102T}}\\ 
HAT-P-4b &  & $0.142_{-0.016}^{+0.014}$  & $0.122_{-0.014}^{+0.012}$  &  &  & {\citet{2013ApJ...770..102T}}\\ 
HAT-P-6b &  & $0.117_{-0.008}^{+0.008}$  & $0.106_{-0.006}^{+0.006}$  &  &  & {\citet{2012ApJ...746..111T}}\\ 
HAT-P-7b &  & $0.098_{-0.017}^{+0.017}$  & $0.159_{-0.022}^{+0.022}$  & $0.245_{-0.031}^{+0.031}$  & $0.225_{-0.052}^{+0.052}$  & {\citet{2010ApJ...710...97C}}\\ 
HAT-P-8b &  & $0.131_{-0.01}^{+0.007}$  & $0.111_{-0.007}^{+0.008}$  &  &  & {\citet{2012ApJ...746..111T}}\\ 
HAT-P-12b &  & $<0.042$  & $<0.085$  &  &  & {\citet{2013ApJ...770..102T}}\\ 
HAT-P-23b & $0.234_{-0.046}^{+0.046}$  & $0.248_{-0.019}^{+0.019}$  & $0.309_{-0.026}^{+0.026}$  &  &  & {\citet{2014ApJ...781..109O}}\\ 
HAT-P-32b & $0.178_{-0.057}^{+0.057}$  & $0.364_{-0.016}^{+0.016}$  & $0.438_{-0.02}^{+0.02}$  &  &  & {\citet{2014ApJ...796..115Z}}\\ 
HD149026b &  & $0.04_{-0.003}^{+0.003}$  & $0.034_{-0.006}^{+0.006}$  & $0.044_{-0.01}^{+0.01}$  & $0.0411_{-0.0076}^{+0.0076}$ $0.052_{-0.006}^{+0.006}$  & {\citet{2012ApJ...754..136S,2009ApJ...703..769K}}\\ 
HD189733b &  & $0.256_{-0.014}^{+0.014}$ $0.1466_{-0.004}^{+0.004}$  & $0.214_{-0.02}^{+0.02}$ $0.1787_{-0.0038}^{+0.0038}$  & $0.31_{-0.034}^{+0.034}$  & $0.3381_{-0.0055}^{+0.0055}$ $0.391_{-0.022}^{+0.022}$  & {\citet{2008ApJ...686.1341C,2012ApJ...754...22K,2007Natur.447..183K}}\\ 
HD209458b &  & $0.094_{-0.009}^{+0.009}$  & $0.213_{-0.015}^{+0.015}$ $0.1391_{-0.0069}^{+0.0072}$  & $0.301_{-0.043}^{+0.043}$  & $0.24_{-0.026}^{+0.026}$  & {\citet{2008ApJ...673..526K,2014ApJ...790...53Z}}\\ 
KELT-1b & $0.160_{-0.020}^{+0.018}$ & $0.195_{-0.010}^{+0.010}$ & $0.200_{-0.012}^{+0.012}$ & & &{\citet{2015ApJ...802...28C,2014ApJ...783..112B}}\\
Kepler-5b &  & $0.103_{-0.017}^{+0.017}$  & $0.107_{-0.015}^{+0.015}$  &  &  & {\citet{2011ApJS..197...11D}}\\ 
Kepler-6b &  & $0.069_{-0.027}^{+0.027}$  & $0.151_{-0.019}^{+0.019}$  &  &  & {\citet{2011ApJS..197...11D}}\\ 
Kepler-12b &  & $0.137_{-0.02}^{+0.02}$  & $0.116_{-0.031}^{+0.031}$  &  &  & {\citet{2011ApJS..197....9F}}\\ 
Kepler-13b & $0.122_{-0.051}^{+0.051}$  & $0.156_{-0.031}^{+0.031}$  & $0.222_{-0.023}^{+0.023}$  &  &  & {\citet{2014ApJ...788...92S}}\\ 
Kepler-17b &  & $0.25_{-0.03}^{+0.03}$  & $0.31_{-0.035}^{+0.035}$  &  &  & {\citet{2011ApJS..197...14D}}\\ 
OGLE-TR-113b & $0.17_{-0.05}^{+0.05}$  &  &  &  &  & {\citet{2007MNRAS.375..307S}}\\ 
Qatar-1b & $0.136_{-0.034}^{+0.034}$  &  &  &  &  & {\citet{2015ApJ...802...28C}}\\ 
TrES-1b &  & $0.083_{-0.024}^{+0.024}$  & $0.066_{-0.013}^{+0.013}$ $0.094_{-0.024}^{+0.024}$  & $0.152_{-0.042}^{+0.042}$  & $0.225_{-0.036}^{+0.036}$ $0.213_{-0.042}^{+0.042}$  & {\citet{2014ApJ...797...42C,2005ApJ...626..523C}}\\ 
TrES-2b & $0.062_{-0.011}^{+0.013}$  & $0.127_{-0.021}^{+0.021}$  & $0.23_{-0.024}^{+0.024}$  & $0.199_{-0.054}^{+0.054}$  & $0.359_{-0.06}^{+0.06}$  & {\citet{2010ApJ...717.1084C,2010ApJ...710.1551O}}\\ 
TrES-3b & $0.241_{-0.043}^{+0.043}$ $0.133_{-0.016}^{+0.018}$  & $0.346_{-0.035}^{+0.035}$  & $0.372_{-0.054}^{+0.054}$  & $0.449_{-0.097}^{+0.097}$  & $0.475_{-0.046}^{+0.046}$  & {\citet{2009A&amp;A...493L..35D,2010ApJ...718..920C,2010ApJ...711..374F}}\\ 
TrES-4b &  & $0.137_{-0.011}^{+0.011}$  & $0.148_{-0.016}^{+0.016}$  & $0.261_{-0.059}^{+0.059}$  & $0.318_{-0.044}^{+0.044}$  & {\citet{2009ApJ...691..866K}}\\ 
WASP-1b &  & $0.184_{-0.016}^{+0.016}$  & $0.217_{-0.017}^{+0.017}$  & $0.274_{-0.058}^{+0.058}$  & $0.474_{-0.046}^{+0.046}$  & {\citet{2010arXiv1004.0836W}}\\ 
WASP-2b & $<0.07$  & $0.083_{-0.035}^{+0.035}$  & $0.169_{-0.017}^{+0.017}$  & $0.192_{-0.077}^{+0.077}$  & $0.285_{-0.059}^{+0.059}$  & {\citet{2010arXiv1004.0836W}}; This Work\\ 
WASP-3b & $0.181_{-0.02}^{+0.02}$ $0.193_{-0.014}^{+0.014}$  & $0.209_{-0.028}^{+0.04}$  & $0.282_{-0.012}^{+0.012}$  &  & $0.328_{-0.055}^{+0.086}$  & {\citet{2012ApJ...748L...8Z,2015ApJ...802...28C,2014MNRAS.441.3666R}}\\ 
WASP-4b & $0.185_{-0.013}^{+0.014}$ $0.16_{-0.04}^{+0.04}$  & $0.319_{-0.031}^{+0.031}$  & $0.343_{-0.027}^{+0.027}$  &  &  & {\citet{2011A&amp;A...530A...5C,2011ApJ...727...23B}}; This Work\\ 
WASP-5b & $0.269_{-0.062}^{+0.062}$ $0.20_{-0.02}^{+0.02}$  & $0.197_{-0.028}^{+0.028}$  & $0.237_{-0.024}^{+0.024}$  &  &  & {\citet{2014A&amp;A...564A...6C,2013ApJ...773..124B}}; This Work\\ 
WASP-10b & $0.137_{-0.019}^{+0.013}$  &  &  &  &  & {\citet{2015A&amp;A...574A.103C}}\\ 
WASP-12b & $0.299_{-0.065}^{+0.065}$ $0.296_{-0.014}^{+0.014}$  & $0.421_{-0.011}^{+0.011}$  & $0.428_{-0.012}^{+0.012}$  & $0.696_{-0.06}^{+0.06}$  & $0.696_{-0.096}^{+0.096}$  & {\citet{2012ApJ...744..122Z,2015ApJ...802...28C,2014ApJ...791...36S}}\\ 
WASP-14b &  & $0.224_{-0.19}^{+0.01}$  & $0.224_{-0.018}^{+0.018}$  &  & $0.181_{-0.022}^{+0.022}$  & {\citet{2013ApJ...779....5B}}\\ 
WASP-17b &  &  & $0.229_{-0.013}^{+0.013}$  &  & $0.237_{-0.039}^{+0.039}$  & {\citet{2011MNRAS.416.2108A}}\\ 
WASP-18b & $0.13_{-0.03}^{+0.03}$  & $0.3_{-0.02}^{+0.02}$ $0.304_{-0.019}^{+0.019}$  & $0.37_{-0.03}^{+0.03}$ $0.379_{-0.015}^{+0.015}$  & $0.37_{-0.03}^{+0.03}$  & $0.41_{-0.02}^{+0.02}$  & {\citet{2011ApJ...742...35N,2013MNRAS.428.2645M}}; This Work\\ 
WASP-19b & $0.366_{-0.072}^{+0.072}$ $0.287_{-0.02}^{+0.02}$  & $0.483_{-0.025}^{+0.025}$  & $0.572_{-0.03}^{+0.03}$  & $0.65_{-0.11}^{+0.11}$  & $0.73_{-0.12}^{+0.12}$  & {\citet{2010MNRAS.404L.114G,2014MNRAS.445.2746Z,2013MNRAS.430.3422A}}\\ 
WASP-24b &  & $0.159_{-0.013}^{+0.013}$  & $0.202_{-0.018}^{+0.018}$  &  &  & {\citet{2012A&amp;A...545A..93S}}\\ 
WASP-33b & $0.27_{-0.04}^{+0.04}$ $0.244_{-0.02}^{+0.027}$  & $0.26_{-0.05}^{+0.05}$  & $0.41_{-0.02}^{+0.02}$  &  &  & {\citet{2012ApJ...754..106D,2013A&amp;A...550A..54D}}\\ 
WASP-36b & $0.13_{-0.04}^{+0.04}$  &  &  &  &  & This Work\\ 
WASP-43b & $0.194_{-0.029}^{+0.029}$ $0.197_{-0.042}^{+0.042}$ $0.181_{-0.027}^{+0.027}$  & $0.347_{-0.013}^{+0.013}$  & $0.382_{-0.015}^{+0.015}$  &  &  & {\citet{2013ApJ...770...70W,2014A&amp;A...563A..40C,2014MNRAS.445.2746Z,2014ApJ...781..116B}}\\ 
WASP-46b & $0.253_{-0.06}^{+0.063}$ $0.26_{-0.04}^{+0.04}$  &  &  &  &  & {\citet{2014A&amp;A...567A...8C}}; This Work\\ 
WASP-48b & $0.109_{-0.027}^{+0.027}$  & $0.176_{-0.013}^{+0.013}$  & $0.214_{-0.02}^{+0.02}$  &  &  & {\citet{2014ApJ...781..109O}}\\ 
WASP-76b & $<0.3$  &  &  &  &  & This Work\\ 
WASP-80b &  & $0.0455_{-0.01}^{+0.01}$  & $0.0944_{-0.0065}^{+0.0064}$  &  &  & {\citet{2015arXiv150308152T}}\\ 
XO-1b &  & $0.086_{-0.007}^{+0.007}$  & $0.122_{-0.009}^{+0.009}$  & $0.261_{-0.031}^{+0.031}$  & $0.21_{-0.029}^{+0.029}$  & {\citet{2008ApJ...684.1427M}}\\ 
XO-2b &  & $0.081_{-0.017}^{+0.017}$  & $0.098_{-0.02}^{+0.02}$  & $0.167_{-0.036}^{+0.036}$  & $0.133_{-0.049}^{+0.049}$  & {\citet{2009ApJ...701..514M}}\\ 
XO-3b &  & $0.101_{-0.004}^{+0.004}$  & $0.143_{-0.006}^{+0.006}$  & $0.134_{-0.049}^{+0.049}$  & $0.15_{-0.036}^{+0.036}$  & {\citet{2010ApJ...711..111M}}\\ 
XO-4b &  & $0.056_{-0.006}^{+0.012}$  & $0.135_{-0.007}^{+0.01}$  &  &  & {\citet{2012ApJ...746..111T}}\\ 

\end{longtable}
}
\twocolumn

\bibliographystyle{mn2e}

\bibliography{mybib_bdfound.bib}


\label{lastpage}

\end{document}